\documentclass[12pt,aps,prb,groupedaddress,amsmath,amssymb]{revtex4-1}

\usepackage{graphicx,subfigure}
\usepackage{dcolumn}
\usepackage{bm}
\usepackage{epsf,color}
\usepackage{fancyhdr}
\usepackage{amsfonts}
\usepackage{float, rotating, rotfloat}
\usepackage[section]{placeins} 
\usepackage{gensymb} 
\usepackage{setspace}
\usepackage{amsmath}

\DeclareMathOperator{\sgn}{sgn}

\usepackage{algcompatible, algpseudocode}
\usepackage[floatrow]{trivfloat}
\trivfloat{algorithm}

\usepackage{etoolbox}
\makeatletter
\preto{\@verbatim}{\topsep=0pt \partopsep=0pt }
\makeatother

\input vmc.def
\input dmc.def
\input projectors.def

\def\psii{\psi_{i}}
\def\psij{\psi_{j}}
\def\psiij{\psi_{ij}}
\def\psisq{\psi^2}

\def\psir1pN{\psi({\bf r}'_1,{\bf r}_2,...,{\bf r}_N)}

\def\Psir1pN{\Psi({\bf r}'_1,{\bf r}_2,...,{\bf r}_N)}
\def\PsiTroneN{\PsiT({\bf r}_1,{\bf r}_2,...,{\bf r}_N)}
\def\PsiTr1pN{\PsiT({\bf r}'_1,{\bf r}_2,...,{\bf r}_N)}

\def\EL{E_{\rm L}}

\def\rvec{{\bf r}}

\def\Rvec{{\bf R}}

\def\Ylmoone{Y_{lm}(\Omega_1)}

\def\Ylmoonepconj{Y^{*}_{lm}(\Omega_1')}

\def\X2bar{\overline{X^2}}

\def\Nquad{{N_{\rm quad}}}

\def\beq{\begin{eqnarray}}
\def\eeq{\end{eqnarray}}

\begin{document}

\title{Nonlocal pseudopotentials and time-step errors in diffusion Monte Carlo}

\author{Tyler A. Anderson\footnote{taa65@cornell.edu}}
\author{C. J. Umrigar\footnote{CyrusUmrigar@cornell.edu}}
\affiliation{
Laboratory of Atomic and Solid State Physics,\\
Cornell University, Ithaca, NY 14853.}

\begin{abstract}
We present a version of the T-moves approach for treating nonlocal pseudopotentials in diffusion Monte Carlo which has
much smaller time-step errors than the existing T-moves approaches, while at the same time preserving desirable
features such as the upper-bound property for the energy.
In addition, we modify the reweighting factor of the projector used in diffusion Monte Carlo to reduce the time-step error.
The latter is applicable not only to pseudopotential calculations but to all-electron calculations as well.
\end{abstract}

\maketitle



\def\br{{\bf r}}
\def\lmax{{l_{\rm max}}}
\def\gtNL{\tilde{g}^{\rm NL}}
\def\ftNL{\tilde{f}^{\rm NL}}
\def\tNL{t^{\rm NL}}

\section{Introduction}
\label{sec:intro}

Projector Monte Carlo (PMC) methods employ a stochastic realization of the power method to project onto the ground state of a Hamiltonian.
Repeated application of the exponential projector, $e^{\tau(\ET-\hat{H})}$, where $\hat{H}$ is the
Hamiltonian and $\ET$ an estimate of the ground state energy, suppresses all but the ground state,
allowing one to calculate properties of the ground state without explicit access to the ground state wave function.
In this paper, we will use diffusion Monte Carlo (DMC), which is a PMC method wherein the Monte Carlo
walk is performed in real space.
The exact expression for the exponential projector in real space is unknown, necessitating the use
of approximate projectors which result in a time-step error which vanishes in the small time step limit.
The purpose of this paper is to reduce the time-step error and thereby improve the computational
efficiency of DMC by enabling the use of large time steps, particularly when nonlocal pseudopotentials
are used.

Nonlocal pseudopotentials are commonly used in electronic structure calculations to reduce the number of electronic degrees of freedom, enabling
one to simulate large systems at low computational cost. In diffusion Monte Carlo (DMC), the nonlocal pseudopotential in the Hamiltonian
introduces off-diagonal matrix elements in the Green's function which may be of either sign, leading to a sign problem in addition to the
usual sign problem.  Surmounting this issue requires the use of an approximation in addition to the usual small $\tau$ approximation.


The earliest such approximation, known as the locality approximation, employs the trial wave function to replace the nonlocal pseudopotential
by an effective local potential~\cite{HurChr-JCP-87,HamReyLes-JCP-87,MitShiCep-JCP-91}.  The exact energy is obtained in the limit that the trial wave function is exact.
The locality approximation energy is not guaranteed to be an upper bound to the true energy.
A more important drawback is that locality approximation calculations often display large negative spikes in the energy
(see e.g. Fig. 1 of Ref.~\onlinecite{Cas-PRB-06}).


%
%
%

The T-moves approximations~\cite{Cas-PRB-06,CasMorSorFil-JCP-10} were developed to address both these shortcomings.  However, the original T-moves approximation~\cite{Cas-PRB-06} is not
size consistent in that for finite time steps it reduces to the locality approximation as the system size is increased.
To cure this problem, Casula, Moroni, Sorella and Filippi~\cite{CasMorSorFil-JCP-10} developed two size-consistent versions of the
T-moves algorithm.  In all three versions of the T-moves approximation, only the sign-violating matrix elements of the nonlocal pseudopotential
are replaced by an effective local potential.
However, all three versions of T-moves frequently exhibit a much larger time-step error than the locality approximation.
This has not received as much attention as it deserves because for many systems the larger time-step error is not as apparent in
the total energy as it is in the expectation values of observables that do not commute with the Hamiltonian which have been
studied less than the energy.

In this paper, we make three improvements to DMC methodology.
\begin{enumerate}
\item We modify the reweighting term in the DMC algorithm to reduce the time-step error.
This modification is applicable both to pseudopotential and all-electron calculations.

\item Instead of using a linear approximation~\cite{Cas-PRB-06,CasMorSorFil-JCP-10} for the part of the Green's function coming from the nonlocal potential, we derive
an exact expression.  This has only a very minor effect on the time-step error but it is nevertheless desirable to avoid an unnecessary
approximation.

\item An additional Metropolis-Hastings accept-reject step is introduced after each T-move to ensure that an excellent approximation
to the exact distribution is sampled in the limit that the trial wave function is exact, even for finite time steps.
(We explain why the sampled distribution is not quite exact even after adding the Metropolis-Hastings step in Sec.~\ref{sec:acc_rej}.)
In practice, for approximate trial wave functions, this greatly reduces the time-step errors in the energy and especially other observables.
\end{enumerate}

The rest of the paper is organized as follows.  The three improvements mentioned above are discussed in Sections
~\ref{sec:rewt}, \ref{sec:exact_Green} and \ref{sec:acc_rej} respectively.
Then in Section~\ref{sec:results} we demonstrate the improvements by showing plots of the total energy and kinetic energy versus time step
for the C and Cr atoms, H$_2$O, C$_4$H$_6$ (butadiene) and a Si$_{15}$ cluster.
We also demonstrate that the T-moves approximation is more efficient than the locality approximation despite a slightly
higher computational cost per Monte Carlo move.
In Section~\ref{sec:outlook} we discuss directions for some future work.

\section{Diffusion Monte Carlo and the reweighting factor of the Green's Function}
\label{sec:dmc_and_rewt}

\subsection{Standard diffusion Monte Carlo}
\label{sec:dmc}

We consider a system containing $N$ electrons at position $\Rvec=\{\rvec_1,\rvec_2,\cdots,\rvec_N\}$, where
we use capital letters for $3N$-dimensional vectors and lowercase letters for $3$-dimensional vectors.
Diffusion Monte Carlo employs the importance-sampled Green's function
\beq
\label{eq:Green}
\Grprtau &=& {\PsiTRp \over \PsiTR} \brakett{\Rvecp}{e^{\tau(\ET-\hat{H})}}{\Rvec}
\eeq
to project onto the mixed distribution $f(\Rvec,t) = \PsiTR\Psi(\Rvec,t)$
using a stochastic implementation of the integral equation
\beq
f(\Rvec',t+\tau) &=& \int d\Rvec \; \Grprtau \; f(\Rvec,t).
\eeq
Here, $\PsiTR$ is a trial wave function, and $\Psi(\Rvec,t)$ for sufficiently large $t$ becomes the lowest energy wave function that has the same nodes as $\PsiT$.
The fixed-node approximation is almost always used, and in that case it is understood that
$\brakett{\Rvecp}{e^{\tau(\ET-\hat{H})}}{\Rvec}$ is evaluated subject to the boundary condition that it vanishes
when $\Rvecp$ is on the nodal surface of $\PsiTR$.
Consequently, the importance-sampled Green's function, $\Grprtau$, vanishes quadratically when $\Rvecp$ is on the nodal surface of $\PsiTR$.

The usual approximate expression for $\Grprtau$ is
\beq
\label{eq:Green2}
\Grprtau &=& {1 \over (2\pi\tau)^{3/2}} e^{-(\Rvecp-\Rvec-\Vvec(\Rvec)\tau)^2 \over 2 \tau} e^{\tau(\ET-(\ELRp+\ELR)/2)},
\eeq
where $\VvecR=\Grad \PsiT / \PsiT$ is the velocity, $\ET$ is an estimate of the ground state energy, and $\ELR = {\hat{H} \PsiR / \PsiR}$.
The stochastic realization of this equation employs {\it walkers} specified by the $3N$ positions of the electrons, $\Rvec$,
and a weight $w$.
Eq.~\ref{eq:Green2} says that a walker at $\Rvec$ drifts a distance $\Vvec(\Rvec)\tau$, then diffuses to a position
$\Rvecp$ obtained by moving each coordinate by $3N$ random numbers drawn from Gaussians of width $\sqrt{\tau}$ centered at the drifted
position $\Rvec+\VvecR\tau$. Finally, the weight of the walker, $w$, is multiplied by the reweighting factor $e^{\tau(\ET-(\ELRp+\ELR)/2)}$.
The DMC energy is then computed as $\langle w \EL \rangle / \langle w \rangle$, where the averages are over
both the walkers in a given Monte Carlo generation, and over a large number of generations.

It is not feasible to use this procedure on systems with nodes in the trial wave function for two reasons.
Firstly, it has an extremely large time-step error. Secondly, it provides an infinite-variance estimator
of the energy since $\ELR$ diverges inversely as the distance to the node and the density of walkers at the node is
finite for any finite $\tau$.
Hence, two kinds of modifications to the above naive algorithm are commonly made.
First, note that even if $\PsiT=\Psiz$, where $\Psiz$ is the true ground state wave function, the above
algorithm fails to sample the exact distribution $\Psizsq$.  This can be remedied~\cite{ReyCepAldLes-JCP-82} by viewing the drift-diffusion
part of $\Grprtau$ in Eq.~\ref{eq:Green2}, $P(\Rvecp,\Rvec) = {1 \over (2\pi\tau)^{3/2}} e^{-(\Rvecp-\Rvec-\Vvec(\Rvec)\tau)^2 \over 2 \tau}$,
as the proposal probability density in a Metropolis-Hastings algorithm.
So, the drift-diffusion step is followed by accepting the proposed move with probability
\beq
\label{eq:acceptance}
p &=& \min\left\{1,{|\PsiTRp|^2 \over |\PsiTR|^2} {P(\Rvec,\Rvecp) \over P(\Rvecp,\Rvec)}\right\}.
\eeq
For systems with more than a few electrons it is more efficient to perform an accept-reject step after moving
each electron, rather than after moving all the electrons of the walker, so that is the version we employ in this work.
Now, the fact that some of the moves are rejected leads to a slower evolution of the system than the one
indicated by time step $\tau$.  Hence, the multiplicative reweighting factor is chosen to be
\beq
\Delta w &=& e^{\taueff(\ET-(\ELRp+\ELR)/2)},
\eeq
where $\tau$ has been replaced by an effective time step,
\beq
\label{eq:taueff}
\taueff &=& \tau {\sum r^2_{\rm acc} \over \sum r^2_{\rm prop}},
\eeq
$\sum r^2_{\rm prop}$ is the sum of the squares of the proposed diffusion steps over the electrons in a walker
and the $\sum r^2_{\rm acc}$ is the same quantity with the sum over only the accepted moves.
Note that if none of the electron moves of a walker are accepted, then that walker does not get reweighted.
Adding this Metropolis-Hastings step to the algorithm makes the sampled distribution go quadratically to zero
at the trial wave function nodes and thereby changes the infinite-variance estimator of the energy into a finite-variance
estimator.

The second set of modifications~\cite{UmrNigRun-JCP-93} to the naive algorithm comes from recognizing that the velocity and the local energy
can have singularities at the nodes of the wave function and at particle coincidences.
Here we focus on the divergences in the velocity and local energy at the nodes.
Near a node, the magnitude of the velocity diverges inversely as the distance to the nodal surface.
Consequently, an electron near a node drifts rapidly away from the node and in so doing the instantaneous
\label{eq:taueff}
velocity drops rapidly.  Hence it is a poor approximation (one that leads to spurious long tails in the
sampled distribution) to assume that the velocity is constant over the time step, $\tau$.
Instead, as discussed in Ref.~\onlinecite{UmrNigRun-JCP-93} and Appendix~\ref{app:integ_vel}, if one
assumes a simple form for the wave function near the node, then the average velocity of an electron over the
time-step, $\tau$, is
\beq
\label{eq:vbar}
\bar{\vvec} &=& \frac{-1+\sqrt{1+2a v^2\,\tau}}{a v^2\,\tau} \vvec \;\;\to\;\; \left\{ \begin{array}{ll}
(1-av^2\tau/2)\vvec &\;\; \mbox{if $v^2\, \tau \ll 1$}\\
\sqrt{{2}/{a\tau}}\; \hat{\vvec} &\;\; \mbox{if $v^2\, \tau \gg 1$}
\end{array} \right. .
\eeq
Of course in the $\tau \to 0$ limit, $\bar{\vvec} =\vvec$.

\subsection{Improved reweighting factor}
\label{sec:rewt}
So far we have presented ingredients of what has come to be the standard algorithm for doing DMC on electronic systems.
We now discuss some ingredients we use in this paper that differ from those in the usual algorithm.
For approximate trial wave functions, the local energy, $\ELR$, diverges to $\pm \infty$ near nodes.
The expression presented in Ref.~\onlinecite{UmrNigRun-JCP-93} to approximate the average of $\ELR$ over $\tau$
is not a very good approximation as clearly demonstrated by Zen et al. in Ref.~\onlinecite{ZenSorGilMicAlf-PRB-16}.
They propose cutting off the local energies according to
\beq
\Sbar(\Rvec) &=& \ET - \Eest + \min\left\{|\Eest-\ELR|,0.2\sqrt{N \over \tau}\right\} \sgn(\Eest-\ELR),
\eeq
where $\Eest$ is the best current estimate of the ground state energy. This has no effect in the $\tau=0$ limit, so when we show energies for their reweighting scheme in
Section~\ref{sec:results} we impose the additional condition that the reweighting factor is
no larger than $10 \sigma_E \tau$, where $\sigma_E$ is the root-mean-square fluctuation of the local energy.

The reweighting factor in Ref.~\onlinecite{UmrNigRun-JCP-93} was chosen by integrating over the drift
path, ignoring completely the diffusion, using the ansatz for the local form of the wave function
described in Appendix~\ref{app:integ_vel}.
Instead, in this paper we present a reweighting factor which was empirically chosen to have a small time-step
error for the carbon atom, which results in smaller time-step
errors than the reweighting factors in either Ref.~\onlinecite{UmrNigRun-JCP-93} or Ref.~\onlinecite{ZenSorGilMicAlf-PRB-16} for all other systems we have tested.

The multiplicative reweighting factor is chosen to be
\beq
\label{eq:rewt}
\Delta w &=& e^{\taueff(\SbarR + \SbarRp)/2},
\eeq
where 
\beq
\label{eq:Sbar}
\SbarR  &=& \ET-\Eest +{E_{\rm cut}(\Rvec) \over 1+(V^2\tau/N)^2}
\;\;\to\;\; \left\{ \begin{array}{ll}
\ET-\ELR  &\;\; \mbox{if $V^2\, \tau \ll N$}\\
\ET-\Eest &\;\; \mbox{if $V^2\, \tau \gg N$}
\end{array} \right. ,
\eeq
where $V^2 = \left| \Grad \PsiT / \PsiT \right|^2$, $N$ is the number of electrons,
\beq
\label{eq:Ecut}
E_{\rm cut}(\Rvec) = \min\left\{|\Eest-\ELR|,10 \sigma_E\right\} \sgn(\Eest-\ELR),
\eeq
and $\sigma_E$ is the current estimate of the root-mean-square fluctuation of $\ELR$.
Note that the factor $1+(V^2\tau/N)^2$ is roughly independent of system size.
In the large $\tau$ limit $\SbarR$ is independent of the starting point $\Rvec$, as it should be.

\section{Exact Nonlocal Green's Function}
\label{sec:exact_Green}

\subsection{Form of pseudopotential}

For a one-body pseudopotential with nonlocal components, the Schr\"odinger equation is
\beq
\sum_{i=1}^N\left(\left(-{1 \over 2} \nabla_i^2 + v_{\rm L}(\rvec_i) + \sum_{j<i}{1 \over r_{ij}} \right) \Psi(\Rvec)
+ \int d\rvecp_i \; v_{\rm NL}(\rvec_i,\rvecp_i) \Psi(\rvec_1,\cdots,\rvecp_i,\cdots,\rvec_N) \right) = E \Psi(\Rvec),
\eeq
where $v_{\rm L}$ and $v_{\rm NL}$ are the local and nonlocal parts of the pseudopotential.
We consider now the effect of the nonlocal potential on electron 1.  (The extension to other
electrons is obvious.)
The pseudopotentials commonly used in electronic structure calculations have only $l$-dependent nonlocality, so
\beq
\int d\rvecp_1 \; v_{\rm NL}(\rvec_1,\rvecp_1) \Psi(\rvecp_1,\rvec_2,\cdots,\rvec_N)
&=& \sum_{l=0}^\lmax v_l(r_1) \sum_{m=-l}^l \Ylmoone \int d \Omega_1' \; \Ylmoonepconj \; \psir1pN,
\label{eq:nonlocpot2}
\eeq
where the integral over $\Omega_1'$ is an integral over the surface of a sphere of radius $r_1$ whose center coincides with the center of the pseudopotential.
The number of nonlocal components, $\lmax+1$, is small, typically between one and three.

\subsection{Nonlocal pseudopotentials in VMC}



The expression in Eq.~\ref{eq:nonlocpot2} can be simplified~\cite{FahyWangLouie90}.
We use the definition of spherical harmonics
\beq
\label{eq:spherical_harmonics}
Y_{lm}(\theta,\phi) = \sqrt{{(2l+1) \over 4 \pi} {(l-m)! \over (l+m)!}}
\; P_l^m(\cos\theta) \; e^{im\phi},
\eeq
where $P_l^m$ are associated Legendre polynomials.
We choose the $z$ axis to be along the vector from the nucleus to electron $1$
so that $\theta_1=0$ and $\cos \theta_1=1$.
For $m \ne 0$, $Y_{lm}(0,0)=P_l^m(1)=0$.
For $m = 0$, $P_l^m \equiv P_l$ where $P_l$ are Legendre polynomials and $P_l(1)=1$, so that $Y_{lm}(0,0)=\sqrt{{(2l+1) \over 4 \pi}}$.
Consequently, the sum over $m$ in Eq.~\ref{eq:nonlocpot2} reduces to just
the $m=0$ term and Eq.~\ref{eq:nonlocpot2} becomes
\beq
\int d\rvec_1' \; v_{\rm NL}(\rvec_1,\rvecp_1) \Psi(\rvecp_1,\rvec_2,\cdots,\rvec_N)
&=& \sum_{l}^\lmax (2l+1) v_l(r_1) \int {d \Omega_1' \over 4 \pi} \;
P_l(\cos\theta_1') \; \Psir1pN.
\label{nonlocpot3}
\eeq
The angular integral is typically evaluated on a spherical quadrature grid, so the contribution
of the nonlocal potential acting on electron 1 to the local energy, $\ELR$, is
\beq
{\int d\rvec_1' \; v_{\rm NL}(\rvec_1,\rvecp_1) \PsiT(\rvecp_1,\rvec_2,\cdots,\rvec_N) \over \PsiTroneN}
&\approx&
 \sum_{l}^\lmax (2l+1) v_l(r_1) \sum_{j=1}^{N_{\rm quad}} \;
w_j P_l(\cos\theta_{1j}') \; {\PsiT({\bf r}'_{1j},{\bf r}_2,...{\bf r}_N) \over \PsiTroneN},
\label{nonlocpot4}
\eeq
where $\sum_{i=1}^{N_{\rm quad}} w_i = 1$.
For the quadrature grids that are the vertices of Platonic solids, the quadrature weights are all
$1/N_{\rm quad}$, but more generally the grid points may have unequal weights.
Note that the potential acting on electron 1 depends on the positions of all the electrons.
The convergence of the integral with respect to the number of quadrature points has been
studied in Ref.~\onlinecite{MitShiCep-JCP-91}. The error goes down a bit faster than $1/N_{\rm quad}^2$.
Note however, that since the contribution of the nonlocal potential to $\ELR$ is linear,
the use of a finite grid does not cause any bias in variational Monte Carlo (VMC).
This is not the case in DMC.

\subsection{Nonlocal pseudopotentials in DMC}

In DMC we need not only the contribution of the nonlocal potential to the local energy, but
also to the Green's function, which now acquires additional nonlocal components.
Separating the local and nonlocal components of the Hamiltonian,
$\hat{H} = \hat{H}_{\rm L} + \hat{v}_{\rm NL}$, the importance sampled Green's function is
\beq
G(\Rvecp,\Rvec,\tau)=\frac{\Psi(\Rvec')}{\Psi(\Rvec)} \langle \Rvecp | {\rm e}^{\tau(\ET - \hat{H}_{L} - \hat{v}_{\rm NL})} | \Rvec \rangle .
\eeq
Using the Suzuki-Trotter expansion for small $\tau$, the Green's function can also be split into parts that use local and nonlocal components of the pseudopotential,
\beq
G(\Rvecp,\Rvec,\tau) &\approx& \int d\Rvecpp \; G_{L}(\Rvecp,\Rvecpp,\tau) \; T(\Rvecpp,\Rvec,\tau),
\eeq
where
\beq
G_{L}(\Rvecp,\Rvec,\tau)
&=& \frac{\Psi(\Rvecp)}{\Psi(\Rvec)} \langle \Rvecp | {\rm e}^{\tau(\ET - \hat{H}_{L})} | \Rvec \rangle \\
&\approx& {1 \over (2\pi\tau)^{3/2}} e^{-(\Rvecp-\Rvec-\Vbarvec(\Rvec)\tau)^2 \over 2 \tau} e^{\taueff(\SbarR + \SbarRp)/2},
\label{eq:local_gf}
\eeq
and
\beq
T(\Rvecp,\Rvec,\tau) = \frac{\Psi(\Rvecp)}{\Psi(\Rvec)} \langle \Rvecp | {\rm e}^{-\tau \hat{v}_{\rm NL}} | \Rvec \rangle.
\label{eq:tmov}
\eeq
In Eq.~\ref{eq:local_gf}, $\Vbarvec = \{ \vbarvec_1,...,\vbarvec_N \}$ is the $3N$-dimensional average velocity using the $3$-dimensional average velocities $\vbarvec$ from Eq.~\ref{eq:vbar} for each electron. $\SbarR, \SbarRp$ are defined in Eqs.~\ref{eq:Sbar} and \ref{eq:Ecut} except that
for now they use the local energy of the local part of the Hamiltonian only.
(Later, we will combine it with a factor coming from $T(\Rvecp,\Rvec,\tau)$ to recover the
$\SbarR, \SbarRp$ for the full Hamiltonian, given in Eqs.~\ref{eq:Sbar} and \ref{eq:Ecut}.)
Finally, $T(\Rvecp,\Rvec,\tau)$ is the eponymous Green's function responsible for nonlocal T-moves.

Since $\hat{v}_{\rm NL} = \sum^{N_{\rm elec}}_{i=1} \hat{v}^{i}_{\rm NL}$ is a one-body
operator, the $N$-electron nonlocal Green's function can be factored into $N$ one-electron nonlocal Green's
functions~\cite{CasMorSorFil-JCP-10}
\beq
\label{eq:tmov_1elec}
T(\Rvecp,\Rvec,\tau) =
\prod^{N_{\rm elec}}_{i=1}
\frac{\Psi(\Rveci')}{\Psi(\Rveci)}
\langle {\bf r}_{i}' | {\rm e}^{-\tau \hat{v}^{i}_{\rm NL}} | {\bf r}_{i} \rangle
\;=\; \prod^{N_{\rm elec}}_{i=1} t(\Rveci',\Rveci,\tau),
\eeq
where $\Rveci' = \left\{\rvec_{1}',...,{\rvec_{i}}',{\rvec_{i+1}},...,\rvec_N\right\}$,
$\Rveci = \left\{\rvec_{1}',...,{\rvec_{i-1}}',{\rvec_{i}},...,\rvec_N\right\}$,
and $\rvec_i$ and $\rvec_i'$ are the positions of the $i$-th electron before and after the T-move, respectively.

Next we derive an explicit expression for ${\rm e}^{-\tau \hat{v}^{i}_{\rm NL}}$.
We write
\beq
\hat{v}^{i}_{\rm NL} = \sum^{\lmax}_{l=0} v_{l}(\hat{r}_{i}) \hat{P}^{i}_{l},
\eeq
where $\hat{P}^{i}_{l}$ projects onto the one-electron subspace with orbital angular momentum $l$,
\beq
\hat{P}^{i}_{l} = \sum^{l}_{m=-l} |lm\rangle_{i}\langle lm|_{i} .
\eeq
Expanding the nonlocal Green's function operator in a Taylor series,
\beq
\label{gfnl1}
{\rm e}^{-\tau \hat{v}^{i}_{\rm NL}}
= {\rm exp}\left( -\tau \sum^{\lmax}_{l=0} v_{l}(\hat{r}_{i}) \hat{P}^{i}_{l}\right) =
\sum^{\infty}_{n=0} \frac{1}{n!} \left( -\tau \sum^{\lmax}_{l=0} v_{l}(\hat{r}_{i}) \hat{P}^{i}_{l} \right)^{n}.
\eeq
Using that $\hat{P}^{i}_{l}$ is a projector,
\beq
\label{projl}
\hat{P}^{i}_{l_1} \hat{P}^{i}_{l_2} = \delta_{l_1,l_2}\hat{P}^{i}_{l_1},
\eeq
it follows that
\beq
\left( -\tau \sum^{\lmax}_{l=0} v_{l}(\hat{r}_{i}) \hat{P}^{i}_{l} \right)^{n} = \sum^{\lmax}_{l=0} \left( -\tau v_{l}(\hat{r}_{i})
\right)^{n} \hat{P}^{i}_{l}.
\eeq
Therefore,
\beq
\sum^{\infty}_{n=0} \frac{1}{n!} \left( -\tau \sum^{\lmax}_{l=0} v_{l}(\hat{r}_{i}) \hat{P}^{i}_{l} \right)^{n} = {\cal I} + \sum^{\infty}_{n=1}
\frac{1}{n!} \sum^{\lmax}_{l=0} \left( -\tau v_{l}(\hat{r}_{i}) \right)^{n} \hat{P}^{i}_{l} =
{\cal I} + \sum^{\lmax}_{l=0} (e^{-\tau v_{l}(\hat{r_{i}})} - 1) \hat{P}^{i}_{l} .
\eeq
In the position basis,
\beq
\langle {\bf r}_i' | \hat{P}^{i}_l | {\bf r}_i \rangle = \sum_{m=-l}^l
Y_{lm}(\theta_i', \phi_i') Y^{*}_{lm}(\theta_i, \phi_i) = \frac{2l+1}{4\pi}
P_l(\cos\Theta_i),
\eeq
where $P_{l}$ are Legendre polynomials and $\Theta_i$ is the angle between ${\bf r}_i'$ and ${\bf r}_i$.
Consequently,
\beq
\langle {\bf r}_i' | {\rm e}^{-\tau v^{i}_{\rm NL}} | {\bf r}_i \rangle = \delta(\br'_i-\br_i)+ \delta(r_i'-r_i)\sum_{l=0}^{\lmax}
\left(e^{-\tau v_l(r_i)}-1\right) \frac{2l+1}{4\pi}
P_l(\cos\Theta_i) .
\eeq
With importance sampling the exact expression for a one-electron T-move becomes
\beq
t(\Rveci',\Rveci,\tau)
&=&
\label{gtNL3}
\delta(\br'_i-\br_i) +
\delta(r'_i-r_i)
\left\{
\frac{\Psi(\Rveci')}{\Psi(\Rveci)}
\sum_{l=0}^\lmax
\left(e^{-\tau v_l(r_i)}-1\right) \frac{2l+1}{4\pi}
P_l(\cos\Theta_i)
\right\} \\
&\equiv&
\label{gtNL4}
\delta(\br'_i-\br_i) + \delta(r'_i-r_i) \tNL(\Rveci',\Rveci) .
\eeq

This can be implemented stochastically by sampling one-electron T-moves from the normalized probability density
\beq
\label{continuous_tmove}
P(\Rveci', \Rveci) = \frac{t(\Rveci',\Rveci)}{\int d {\bf \rvec}_i'' ~ t(\Rveci'',\Rveci)}
\eeq
for each of the electrons
and reweighting the walker by multiplicative factors equal to the denominator of Eq.~\ref{continuous_tmove}.
These reweighting factors combine with the reweighting factors coming from the local part of
the Hamiltonian to recover the usual reweighting factor in Eqs.~\ref{eq:Sbar}, \ref{eq:Ecut} obtained from the full Hamiltonian.
Note that the probability density consists of a continuous distribution on the sphere with $r_i'=r_i$
and a $\delta$-function at $\rveci'=\rveci$.

In practice a randomly oriented grid of points on the sphere and the heat-bath algorithm to select either one of the
grid points or the initial point are used.
Several grids of quadrature points and associated weights, $w_i$ for various numbers of quadrature
points, $\Nquad$, are available in the literature.
We denote the $j^{th}$ 3-dimensional grid point of the $i^{th}$ electron by $\rvec'_{i,j}$.
Let $\Rvec_{i,j}' = \{\rvec_1',...,\rvec_{i-1}', \rvec_{i,j}',\rvec_{i+1},...,\rvec_N\}$.
The probability of a move to $\Rvec_{i,j}'$ is
\beq
P(\Rvec_{i,j}',\Rvec_i) &=& {w_j \tNL(\Rvec_{i,j}',\Rvec_i) \over
1 + \sum_k^\Nquad w_k \tNL(\Rvec_{i,k}',\Rvec_i)},
\label{eq:tmov_prob}
\eeq
and the probability of staying at the current point is
\beq
P(\Rvec_{i},\Rvec_i) &=& {1 \over
1 + \sum_k^\Nquad w_k \tNL(\Rvec_{i,k}',\Rvec_i)}.
\label{eq:tmov_prob_diag}
\eeq

We have so far swept under the rug the fact that the expression in Eq.~\ref{gtNL3} can be negative
because any of the 3 factors $\frac{\Psi(\Rveci')}{\Psi(\Rveci)}$,
$\left(e^{-\tau v_l(r_i)}-1\right)$ and $P_l(\cos\Theta_i)$ can be negative.
This causes a sign problem, in addition to the sign problem already present for local potentials.
There are two approaches that have been used to circumvent this.
In the locality approximation~\cite{HurChr-JCP-87,MitShiCep-JCP-91} all of the T-moves are dropped.
This is equivalent to zeroing out all the off-diagonal elements of the Green's function
and adding them to the diagonal.
Instead, in the various T-moves approximations~\cite{Cas-PRB-06,CasMorSorFil-JCP-10}, the
sign-preserving T-moves are executed and only the sign-violating T-moves are dropped.
This is equivalent to zeroing out the sign-violating off-diagonal elements of the Green's function
and adding them to the diagonal.
If these T-moves are done after doing the drift-diffusion steps for all the electrons of a walker,
then one recovers version 1 of the size-consistent T-moves procedure described in Ref.~\onlinecite{CasMorSorFil-JCP-10}.
The only difference is that Ref.~\onlinecite{CasMorSorFil-JCP-10} makes a linear approximation to
the expression in Eq.~\ref{gtNL3}, but that is in practice of little consequence.
This is the version of Casula et al.'s T-moves that we will compare to when we show
the time-step errors in Sec.~\ref{sec:results}.


\section{T-moves accept-reject step}
\label{sec:acc_rej}

We now present our simple modification of the T-moves algorithms that achieves
a large reduction in the time-step error.
All three T-moves algorithms in the literature fail to sample the exact distribution $\Psizsq$ in
the ideal limit that $\PsiT=\Psiz$.  This is easily cured by introducing a Metropolis-Hastings
accept-reject step after each one-electron T-move.  The acceptance probability is
\beq
\label{eq:acc_tmove_exact}
A(\Rveci',\Rveci) &=& \min\left\{1,{\PsiT^2(\Rveci') \over \PsiT^2(\Rvec_i)} {P(\Rvec_i,\Rveci') \over P(\Rveci',\Rvec_i)}\right\} \\
\label{eq:acc_tmove_exact_simp}
&=& \min\left\{1, {\int d {\Rveci''}~t(\Rveci'', \Rveci) \over
\int d {\Rveci''}~t(\Rveci'', \Rveci')} \right\} .
\eeq
In going from Eq.~\ref{eq:acc_tmove_exact} to Eq.~\ref{eq:acc_tmove_exact_simp}, we have used $\frac{t(\Rveci, \Rveci')}{t(\Rveci', \Rveci)} = \frac{\PsiT^2(\Rveci)}{\PsiT^2(\Rveci')}$. Since we employ a quadrature grid this becomes
\beq
\label{eq:acc_tmove_approx}
A(\Rvec_{i,j}',\Rveci)
&\approx& \min\left\{1, {1 + \sum_k^\Nquad w_k \tNL(\Rvec_{i,k}',\Rvec_i) \over
1 + \sum_k^\Nquad w_k \tNL(\Rvec_{i,k}'',\Rvec_{i,j}')} \right\}.
\eeq

With the addition of this Metropolis-Hastings step, the algorithm samples a very good
approximation to $\Psizsq$ in the $\Psi \to \Psiz$ limit.
The reason it does not do so precisely
\footnote{
The bias can be detected by turning
off the reweighting, in which case the variational energy should be obtained.
For the wave functions and time-steps normally used this bias is tiny.  However, we can easily detect
the greatly increased bias obtained
by using a wave function without a Jastrow factor and by increasing $\tau$ to the unrealistically large value of $\tau=$ 5 Ha$^{-1}$.
Note that to evaluate the numerator and the denominator in Eq.~\ref{eq:acc_tmove_approx} one can use either the
same grid points for the reverse move as for the forward move, or rotate the quadrature grid to
sample a fresh set of grid points for the reverse move.
Empirical tests show no clear pattern as to which gives
a smaller bias.  However, using the same grid points results in some saving in computer time.}
is that we are using a finite quadrature grid
and the average of a nonlinear function of a random variable is not equal to the
nonlinear function of the average of the random variable.
In this case the nonlinear functions are the ratio and the minimum function in the acceptance probability.
To evaluate the numerator and the denominator in Eq.~\ref{eq:acc_tmove_approx} one can use either the
same grid points for the reverse move as for the forward move, or rotate the quadrature grid to
sample a fresh set of grid points.
In Eq.~\ref{eq:acc_tmove_approx}, we denote the grid points used for the reverse move by $\Rvec_{i,k}''$.
In either case the above mentioned bias is present, but its
value is different.  Empirical tests indicate that neither choice always gives a smaller bias than
the other, but choosing the same grid reduces the computer time slightly.

In the small $\tau$ limit, the proposal probability for an off-diagonal T-move is linear in $\tau$,
and the acceptance probability deviates from one as some positive power of $\tau$.
Since previous versions of T-moves not using this accept-reject step are recovered by setting the acceptance probability identically to one,
the total probability to make an off-diagonal T-move is changed by our modification only superlinearly in $\tau$.
Hence, older versions of T-moves and our modified T-moves should have the same $\tau = 0$ limit.

We note that the convergence of the energies of both the locality approximation and the T-moves approximation
is quadratic in the error of the trial wave function.  This has been shown for the locality approximation
by Mitas et al.~\cite{MitShiCep-JCP-91}, but the same proof applies to the T-moves approximations as well.
Let $H$ be the exact
Hamiltonian, and $H_A$ be the approximate Hamiltonian for a given trial wave function $\PsiT$ in either the locality approximation or the T-moves approximation. Let
$E_0$ and $E_A$ be the ground state energies of $H$ and $H_A$ respectively, and $\Psi_0$ and $\Psi_A$ the ground state eigenstates of $H$
and $H_A$ respectively. Then
\beq
\label{mixed_est}
E_{A} = \frac{\langle \Psi_A | H | \PsiT \rangle}{ \langle \Psi_A | \PsiT \rangle}.
\eeq
If either $\Psi_A = \Psi_0$ or $\PsiT = \Psi_0$ then $E_A = E_{0}$.
From this it follows that there can be no terms linear in the error of $\Psi_T$ or $\Psi_A$ alone
in the Taylor expansion of Eq.~\ref{mixed_est} with respect to the wave function error.
The limit $\PsiT \rightarrow \Psi_0$ implies $\Psi_A \rightarrow \Psi_0$.
Hence, improving $\PsiT$ simultaneously improves both the bra and the ket in Eq.~\ref{mixed_est} and the convergence of $E_A$ to $E_0$ is quadratic in the wave function
error.
In explicit Monte Carlo language, improving $\PsiT$ simultaneously improves the quality of the local energy as well as the quality of the
sampled wave function in both approximations.

As discussed in Ref.~\onlinecite{Cas-PRB-06}, the T-moves algorithms give an upper bound to
the true energy at $\tau=0$.  The proof is in Ref.~\onlinecite{Haaf1995} and applies to any
discrete-space projector Monte Carlo method where the sign problem is circumvented by moving
the off-diagonal elements of the projector to the diagonal.
The proof can be extended to show that the T-moves energy is always higher than the
locality-approximation energy.  The proof is in Appendix~\ref{app:energy_bound}.
The locality approximation can give an energy that is either higher or lower than the true energy.
However it seems likely that for most systems it gives an energy closer to the true energy than
do the T-moves approximations, because
the errors due to the approximated terms in the locality approximation are of both signs,
whereas the errors of all the approximated terms in the T-moves approximation are positive.

Omitting details that are equally applicable to systems with a local potential, such as branching of walkers, population control, and corrections for population
control which are discussed in Ref.~\onlinecite{UmrNigRun-JCP-93}, the algorithm consists
of the following steps for each walker.
\begin{enumerate}
\item For each electron in the walker, propose T-moves with the probabilities in Eq.~\ref{eq:tmov_prob}
and accept them with the probabilities in Eq.~\ref{eq:acc_tmove_approx}.
\item For each electron in the walker, perform the drift, diffusion and accept-reject steps as one
would in a DMC run with a local potential, using the average velocity in Eq.~\ref{eq:vbar} for the drift.
\item Calculate the local energy at the current position and use this and the saved local energy from the previous Monte Carlo (MC) step to reweight the walker according to Eq.~\ref{eq:rewt}.
\end{enumerate}
In steps 1 and 3, the integration grid for each electron is rotated independently.
To save some computer time, in step 1, so long as none of the electrons in a walker have had their T-move
accepted, one can compute the T-moves probabilities using the values of the wave function already computed
in step 3 of the previous MC step.
Note that in the T-moves algorithms of Casula et al. it is important that the reweighting should not
be done after a T-moves step because the density of walkers after their T-moves step is finite at
the nodes and this leads to an infinite variance estimator for the energy.
Instead in our algorithm, since each T-move is followed by an accept-reject step, the 3 steps
outlined above can be done in any order.

\section{Results}
\label{sec:results}

\subsection{Time-step errors}

\begin{figure}[htb]
\centering
\subfigure[]{{\includegraphics[width=3.5in,height=2.6in,clip]{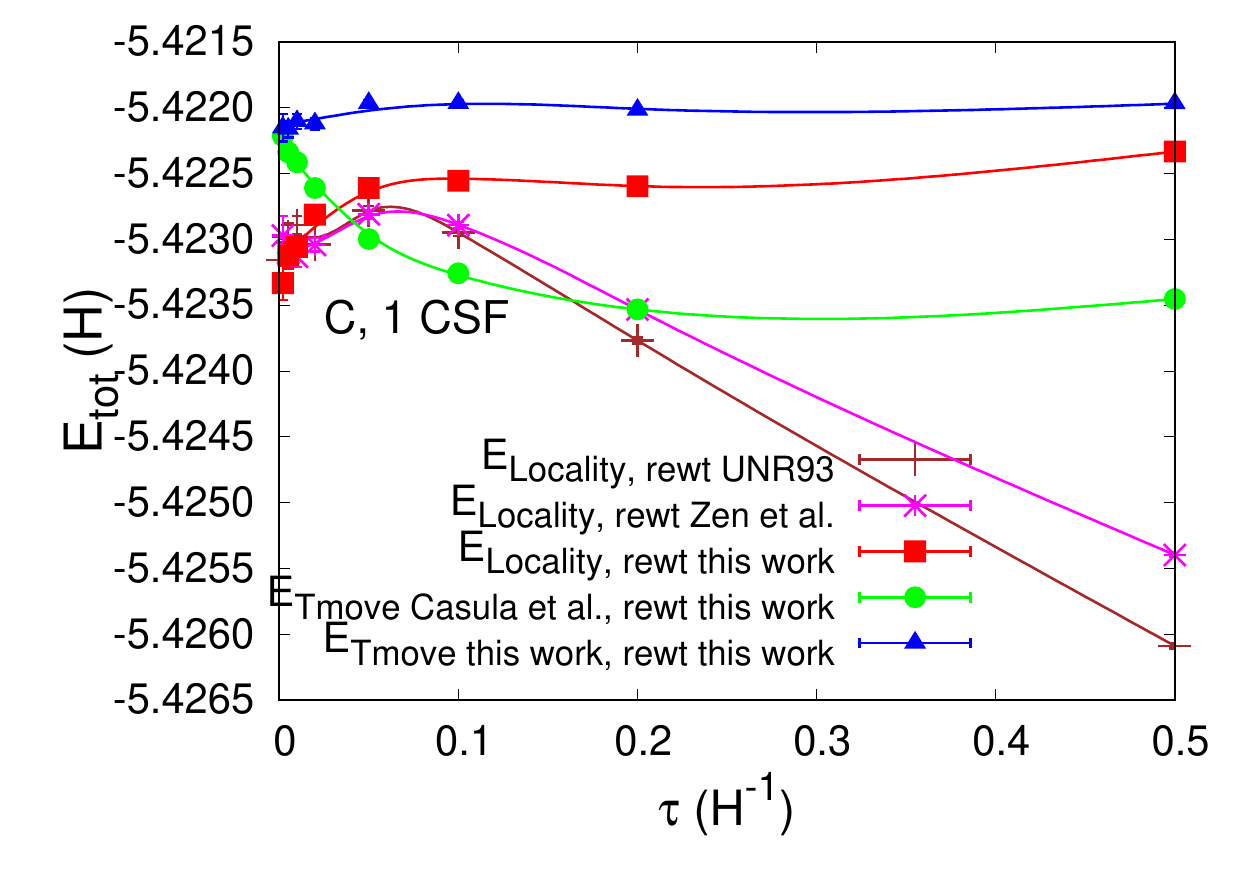}}}\quad
\subfigure[]{{\includegraphics[width=3.5in,height=2.6in,clip]{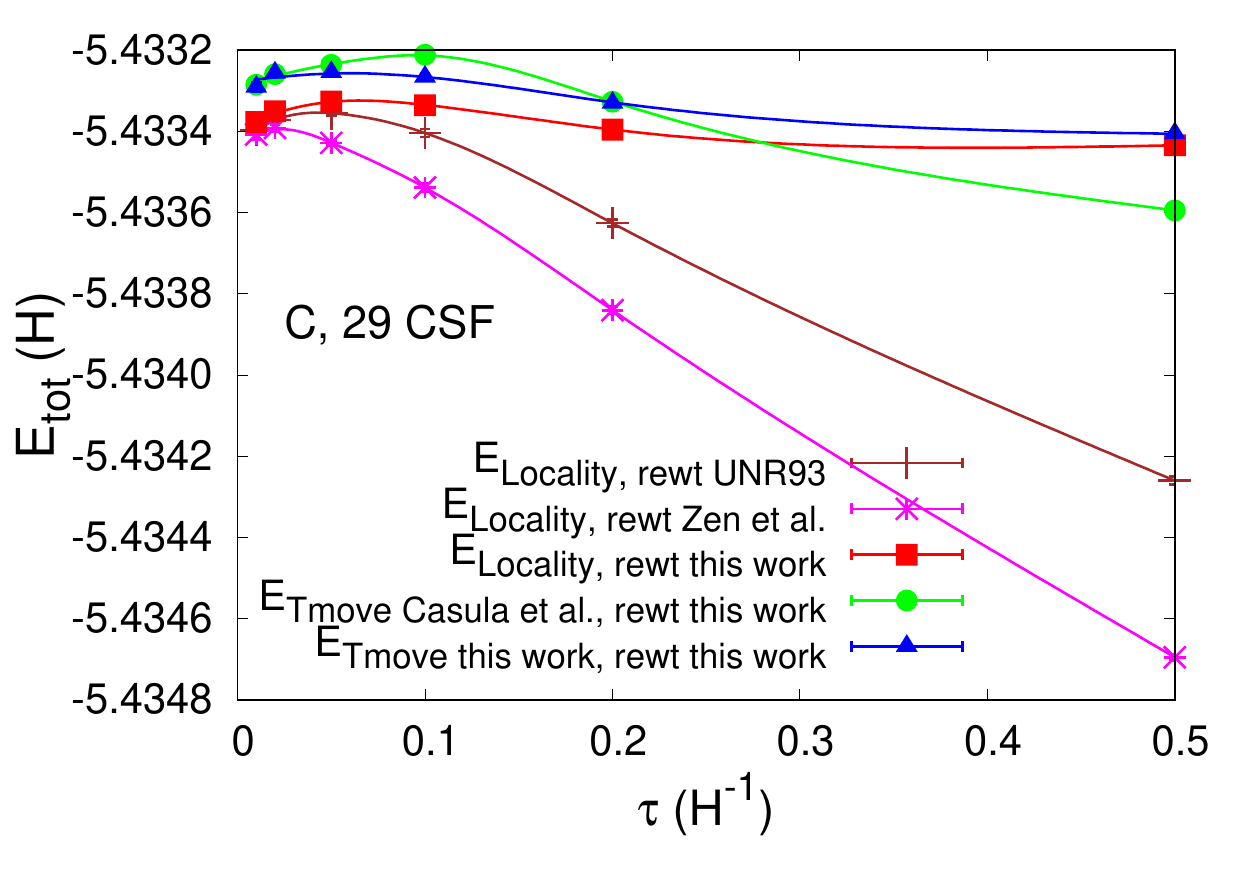}}}\quad\\
\subfigure[]{{\includegraphics[width=3.5in,height=2.6in,clip]{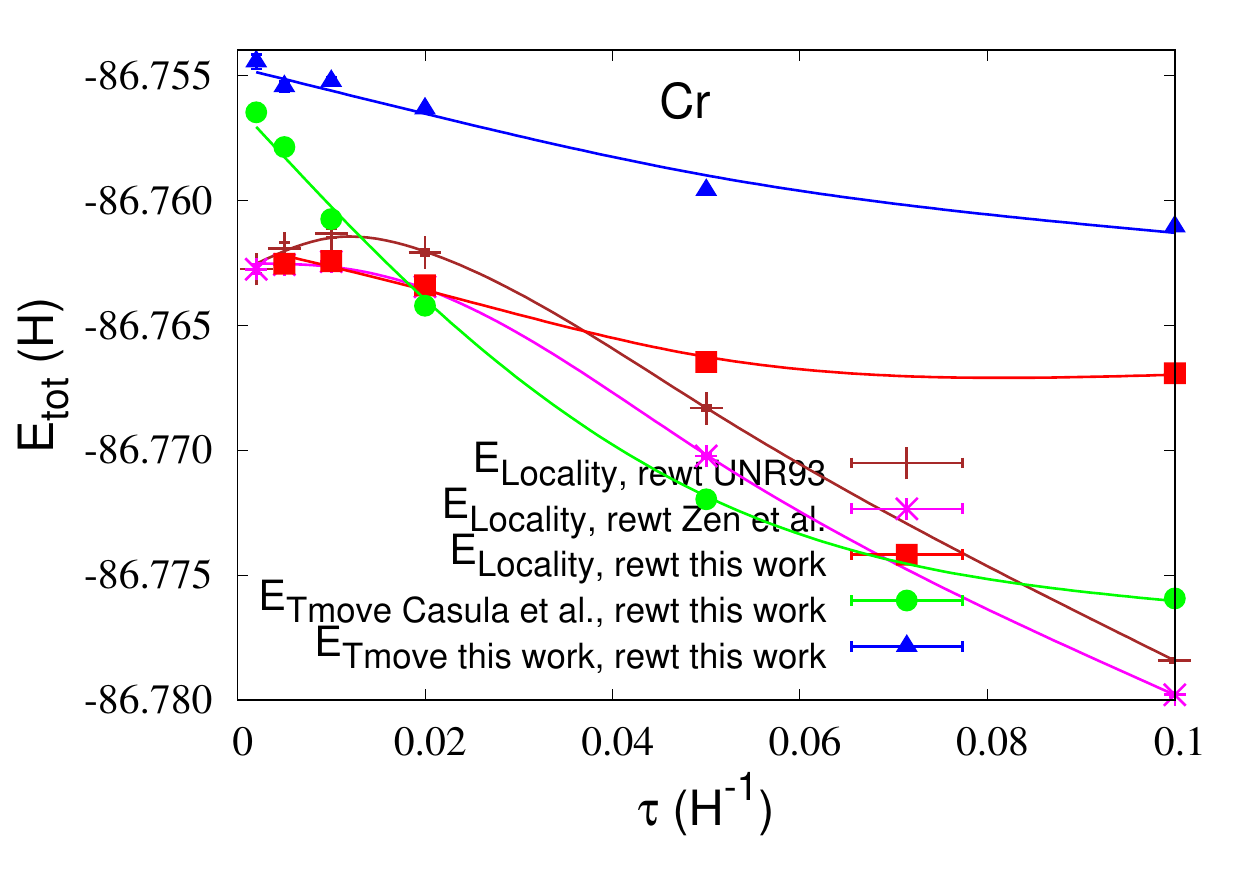}}}\quad
\subfigure[]{{\includegraphics[width=3.5in,height=2.6in,clip]{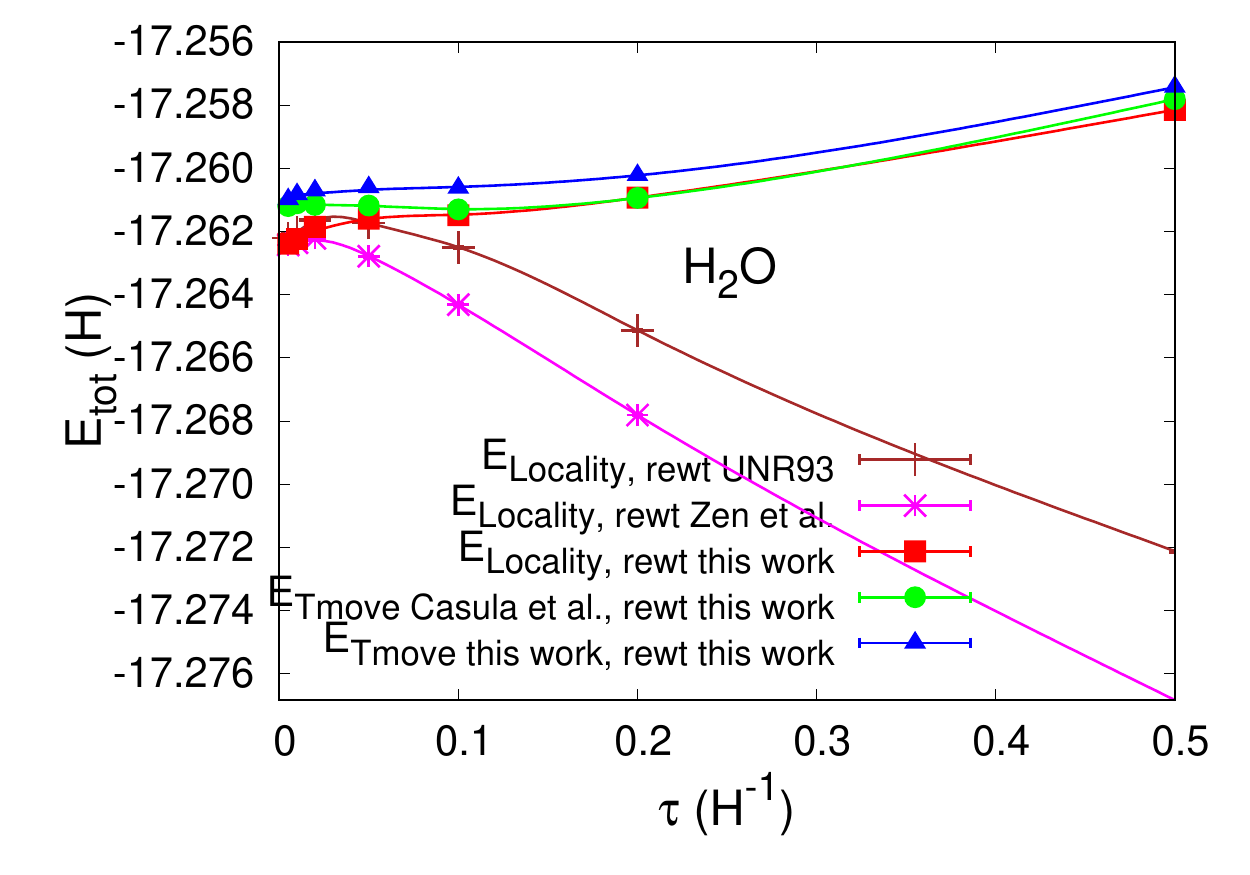}}}\quad
\subfigure[]{{\includegraphics[width=3.5in,height=2.6in,clip]{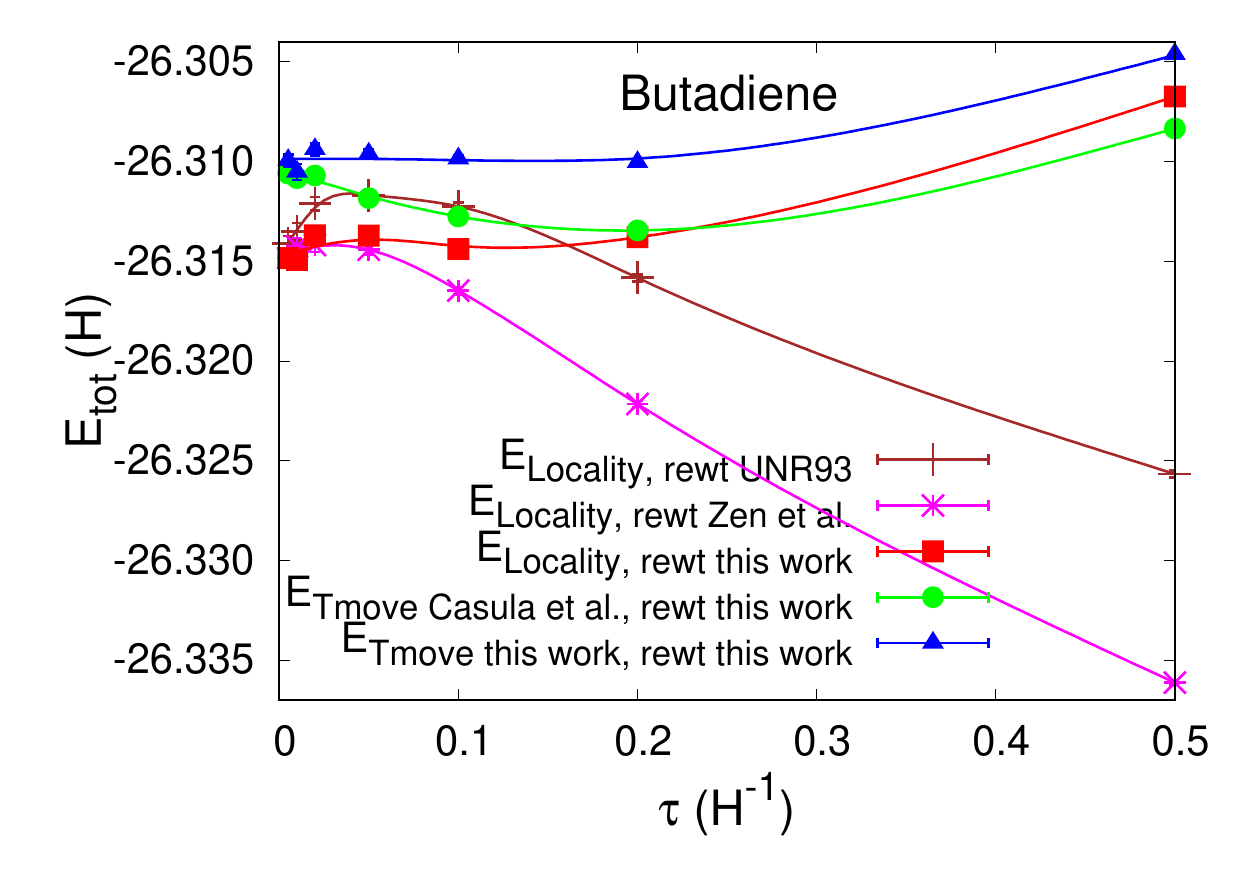}}}\quad
\subfigure[]{{\includegraphics[width=3.5in,height=2.6in,clip]{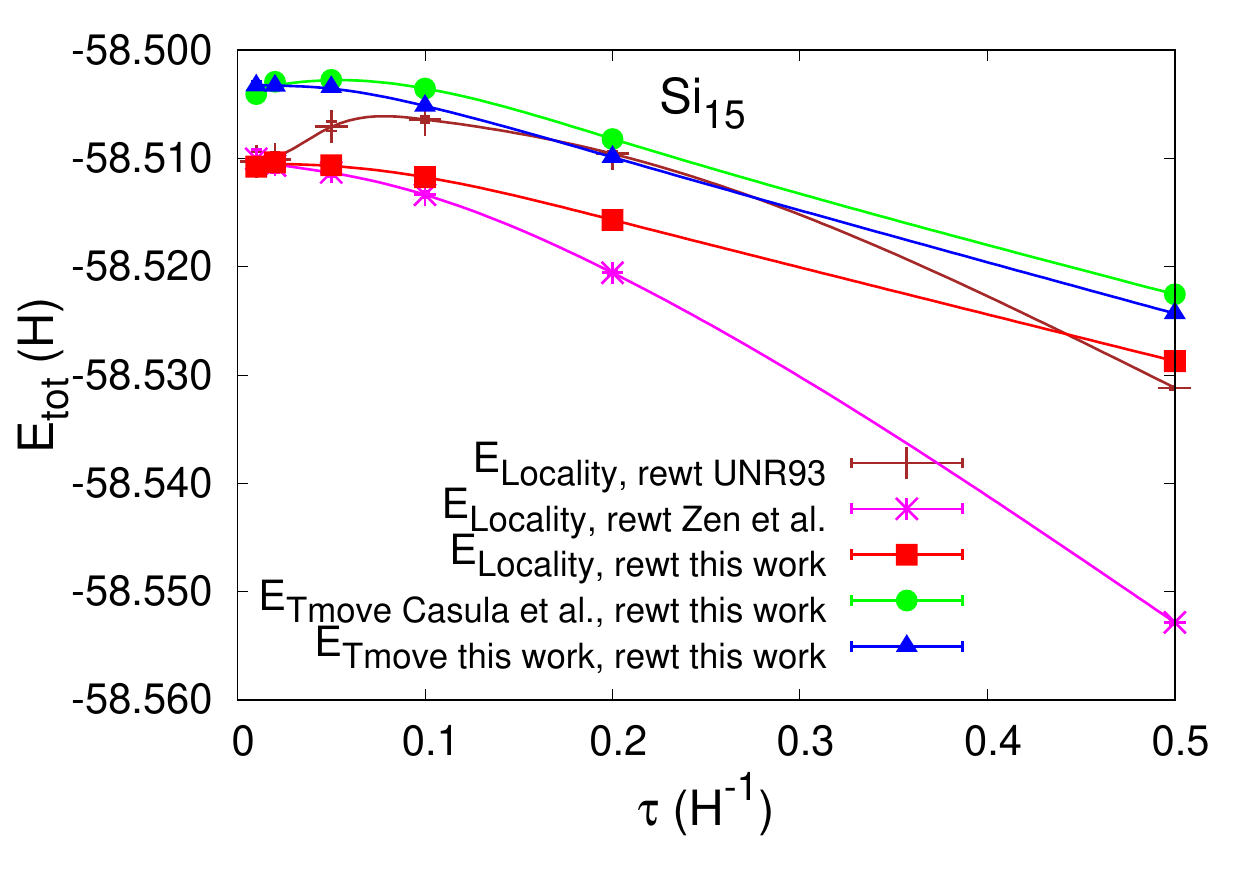}}}\quad
\caption{Time-step errors of the total energy.  The curves labeled ``UNR93" and ``Zen et al." employ the
reweighting factors of Refs.~\onlinecite{UmrNigRun-JCP-93} and \onlinecite{ZenSorGilMicAlf-PRB-16} respectively.
The three locality approximation curves must extrapolate to the same energy at $\tau=0$, as must the two T-moves approximation curves.
The lines are smoothing spline fits to the data.
Both the reweighting factor and the modified T-moves algorithm proposed in this work contribute to
reducing the time-step errors.}
\label{fig:total_energy_timestep}
\end{figure}

\begin{figure}[htb]
\centering
\subfigure[]{{\includegraphics[width=3.5in,height=2.6in,clip]{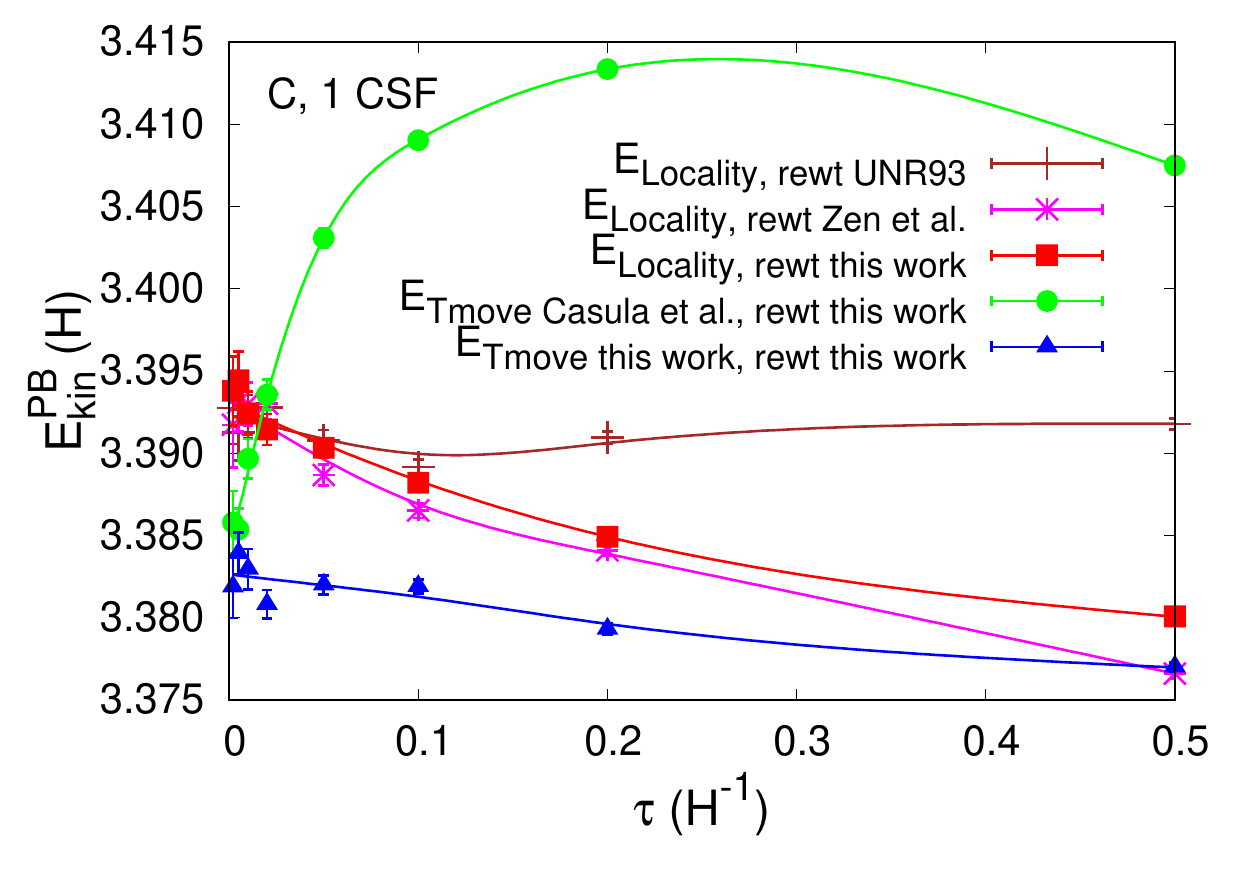}}}\quad
\subfigure[]{{\includegraphics[width=3.5in,height=2.6in,clip]{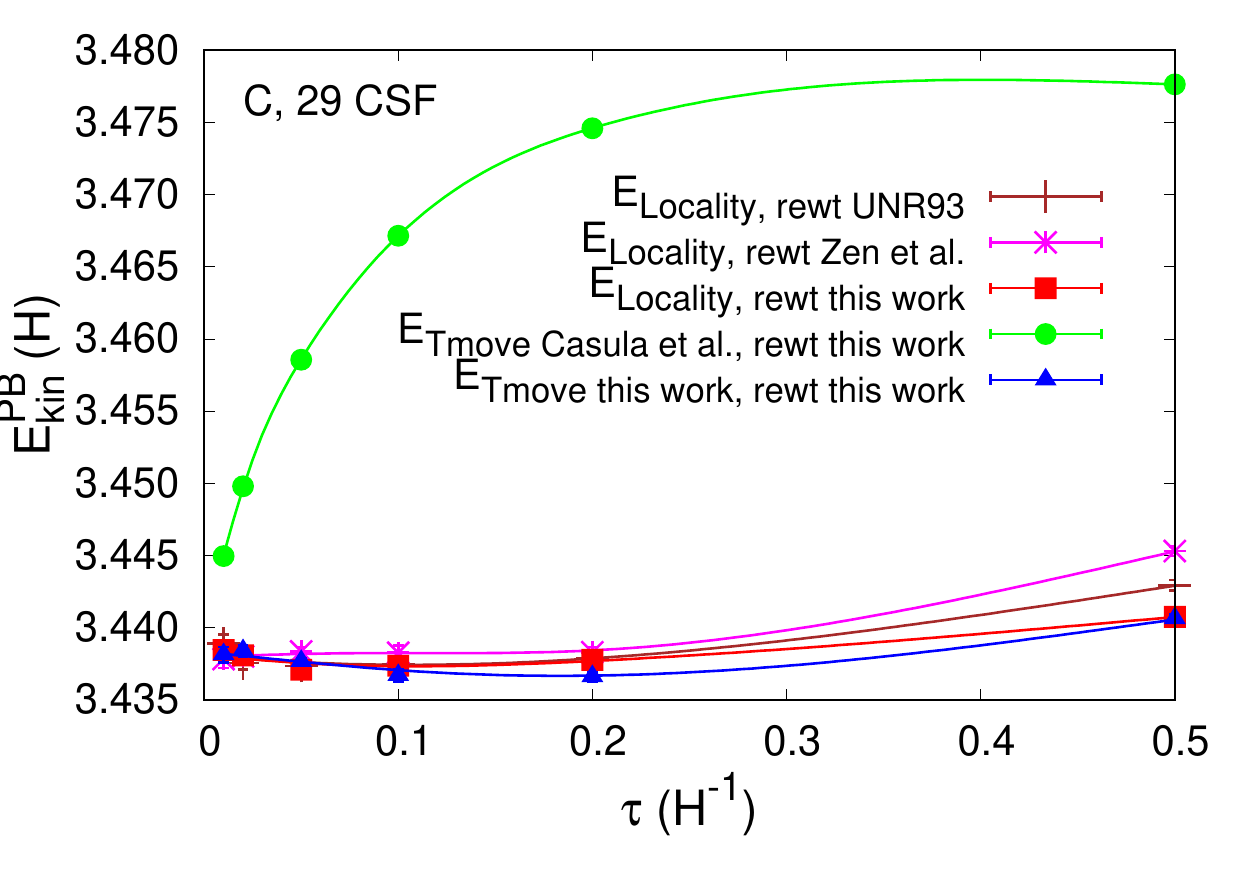}}}\quad\\
\subfigure[]{{\includegraphics[width=3.5in,height=2.6in,clip]{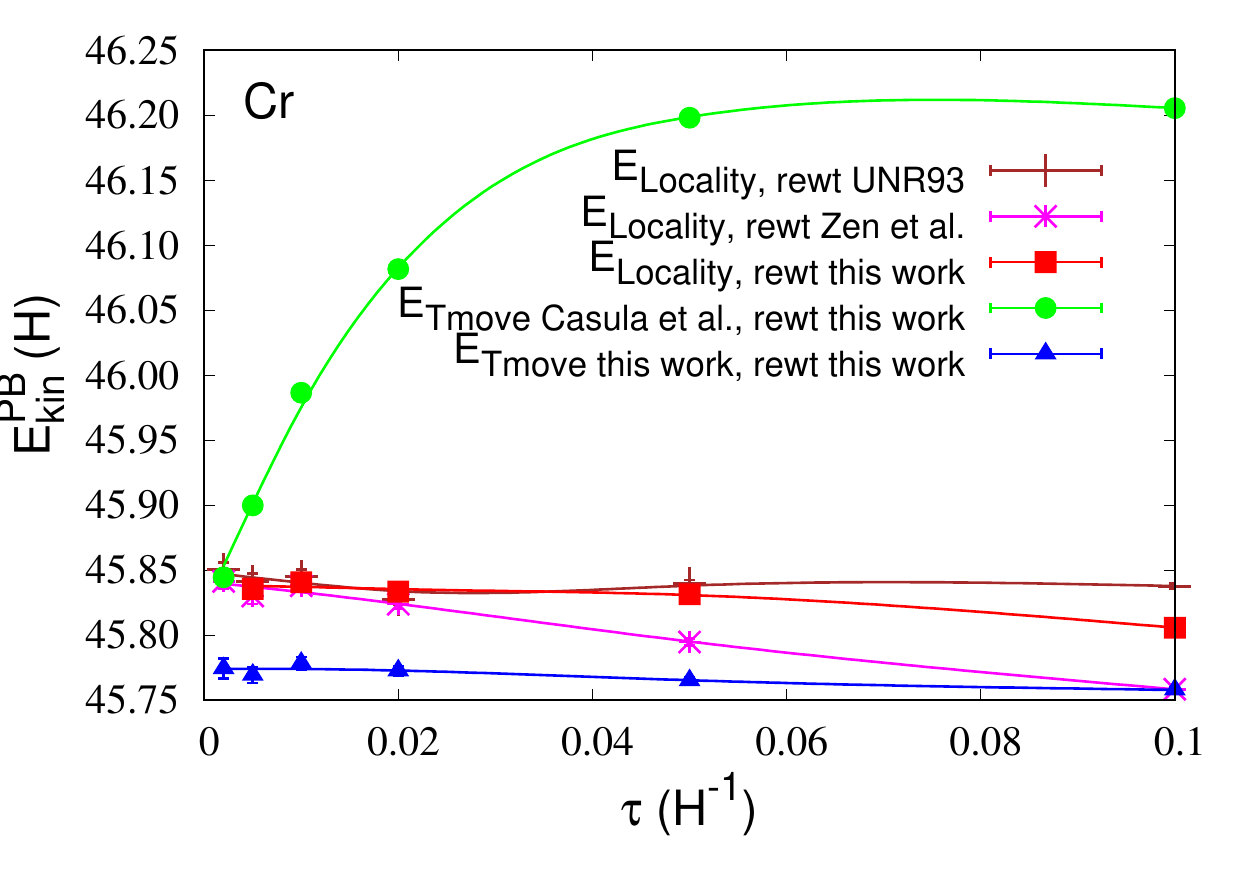}}}\quad
\subfigure[]{{\includegraphics[width=3.5in,height=2.6in,clip]{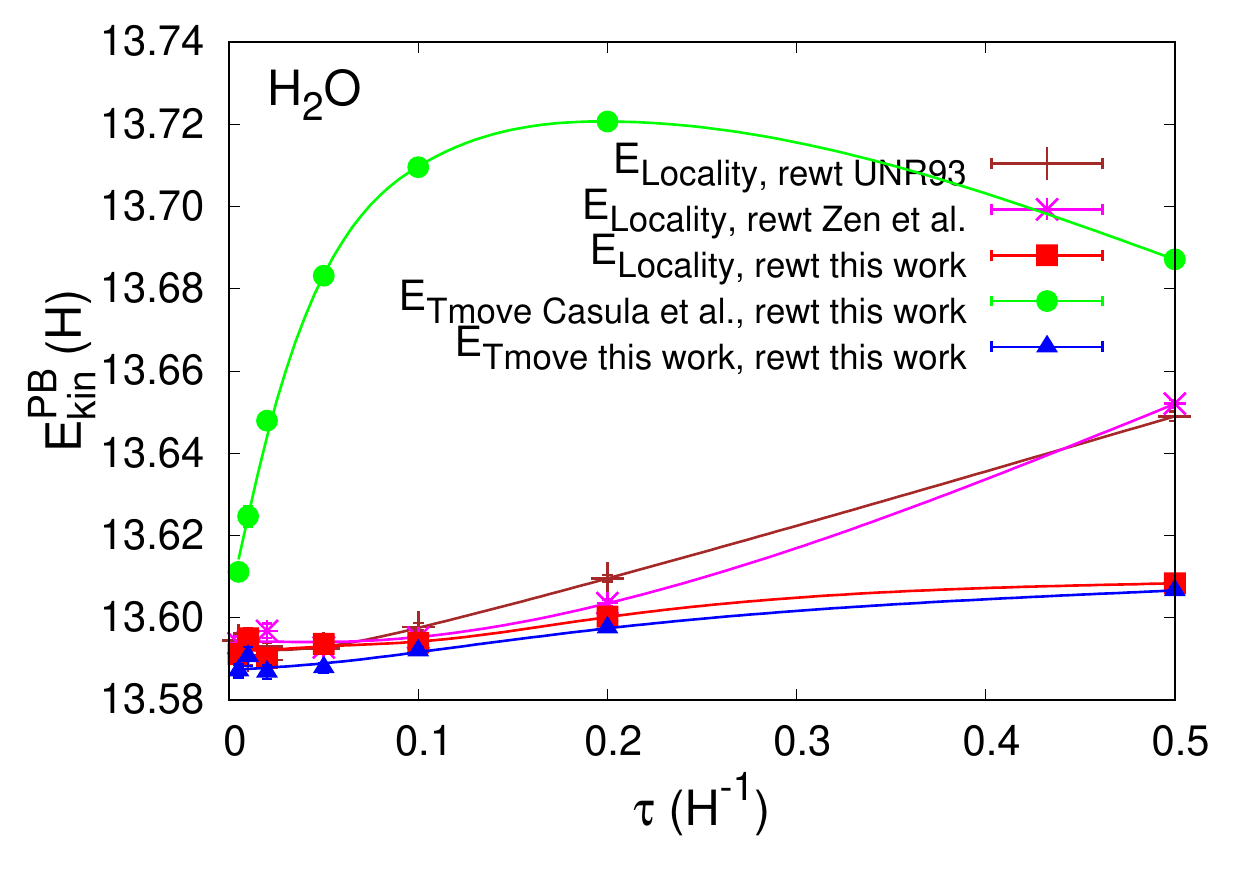}}}\quad\\
\subfigure[]{{\includegraphics[width=3.5in,height=2.4in,clip]{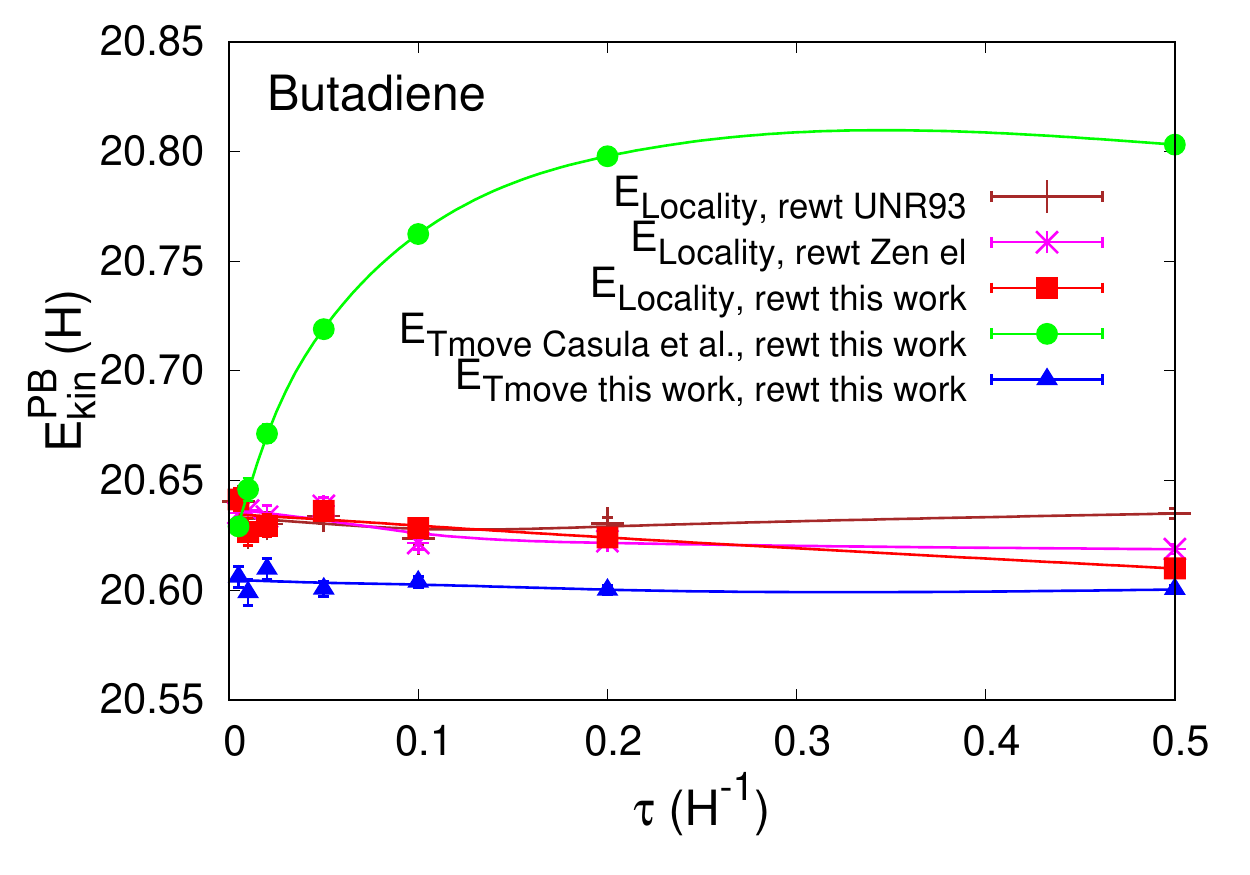}}}\quad
\subfigure[]{{\includegraphics[width=3.5in,height=2.4in,clip]{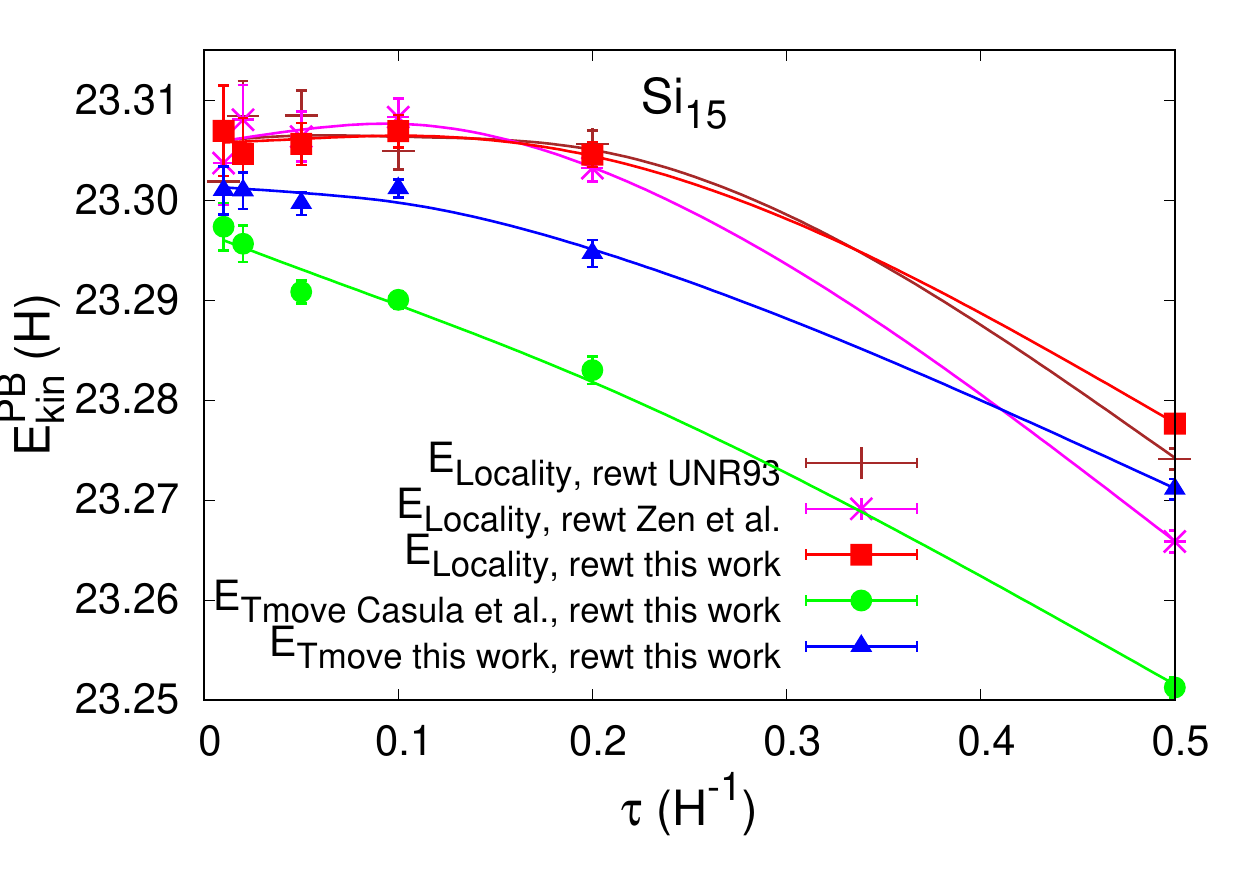}}}\quad
\caption{Same as Fig.~\ref{fig:total_energy_timestep} but for the kinetic energy. Note that the T-moves curves should converge to the same value at $\tau=0$, though this is to some extent obscured by the statistical
errors and the steepness of the kinetic energy curves.}
\label{fig:kinetic_energy_timestep}
\end{figure}

We demonstrate the reduction in the time-step error resulting from the improved reweighting procedure and from
the improved T-moves algorithm by plotting, for two atoms and three molecules, the total energy and the kinetic energy
versus the time step in Figs.~\ref{fig:total_energy_timestep} and \ref{fig:kinetic_energy_timestep} respectively.
The two atoms studied are carbon and chromium.
For carbon, we use the Burkatzki-Filippi-Dolg (BFD) pseudopotential~\cite{BurFilDol-JCP-08},
an atomic-natural-orbital Gauss-Slater (ANO-GS) basis~\cite{PetTouUmr-JCP-11}, and either a single-determinant trial wave function or one with 29 configuration state functions (CSFs).
For chromium, we use the Trail-Needs pseudopotential~\cite{TraNee-JCP-15,TraNee-JCP-17} and a trial wave function with a single CSF.
The three molecular systems are water, butadiene (C$_4$H$_6$) and a Si$_{15}$ cluster, using the BFD pseudpotentials,
\footnote{The hydrogen pseudopotential on the web site, http:\/\/burkatzki.com\/pseudos\/index.2.html, is incorrect, so we use
the corrected pseudopotential communicated to us by Filippi and Dolg.}
the ANO-GS bases, and a single determinant.
We used small basis sets (3-zeta for Cr and Si$_{15}$ and 2-zeta for the other four systems) because DMC has a fairly fast convergence with
basis size, and because for the purposes of this paper achieving convergence with respect to the basis size is not important.
The Jastrow, orbital and CSF parameters (if any) were simultaneously optimized using the
{\it linear method}~\cite{TouUmr-JCP-07,TouUmr-JCP-08}.

\begin{figure}[t!]
\centering
\subfigure[]{{\includegraphics[width=3.5in,height=2.6in,clip]{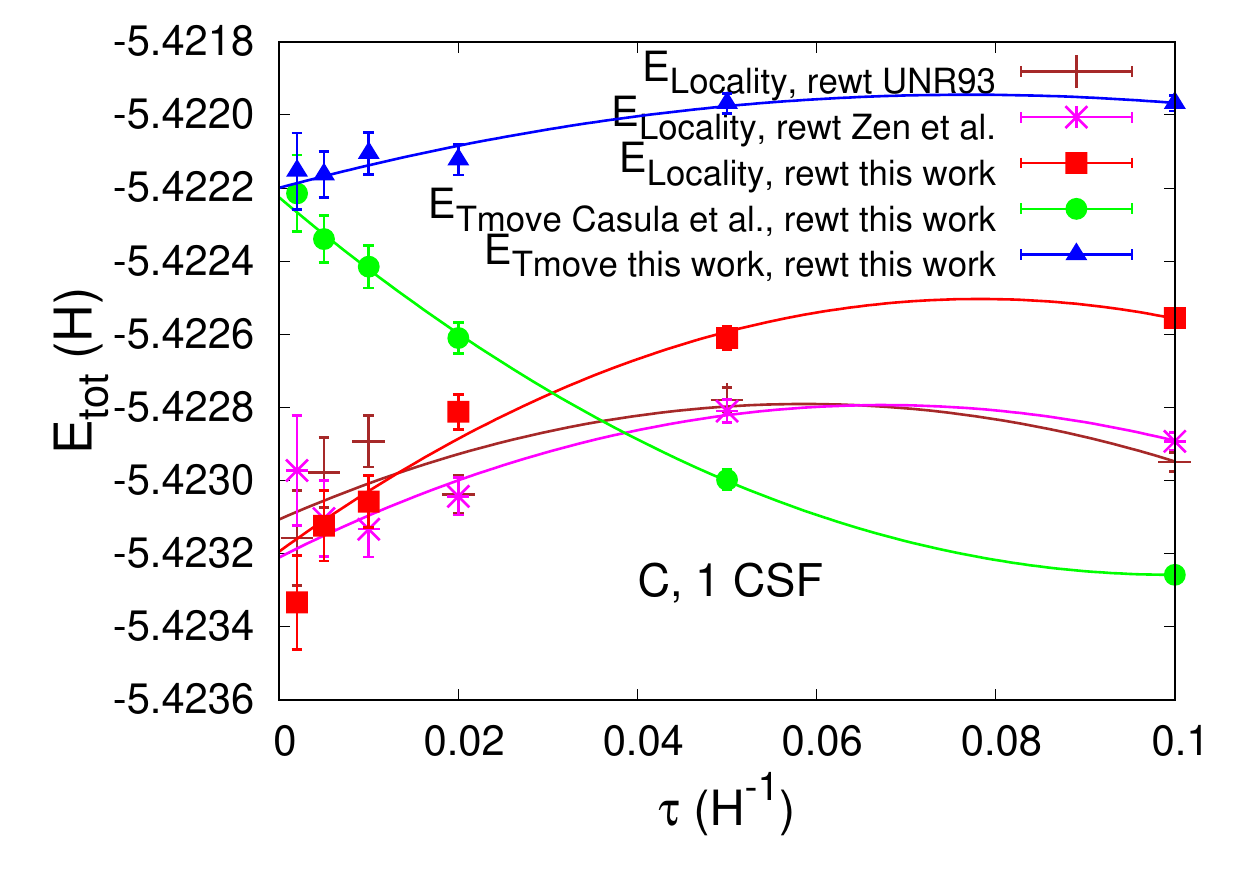}}}\quad
\subfigure[]{{\includegraphics[width=3.5in,height=2.6in,clip]{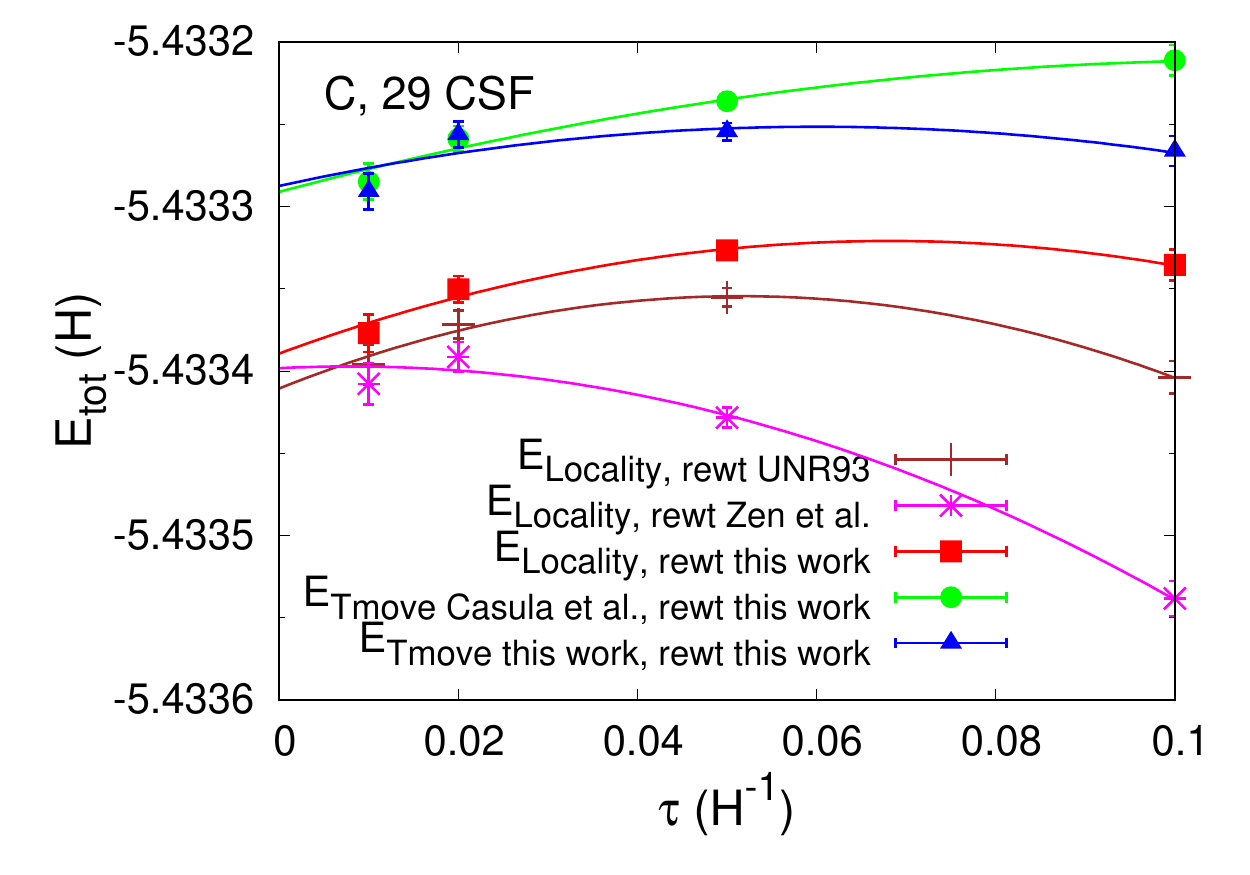}}}\quad\\
\subfigure[]{{\includegraphics[width=3.5in,height=2.6in,clip]{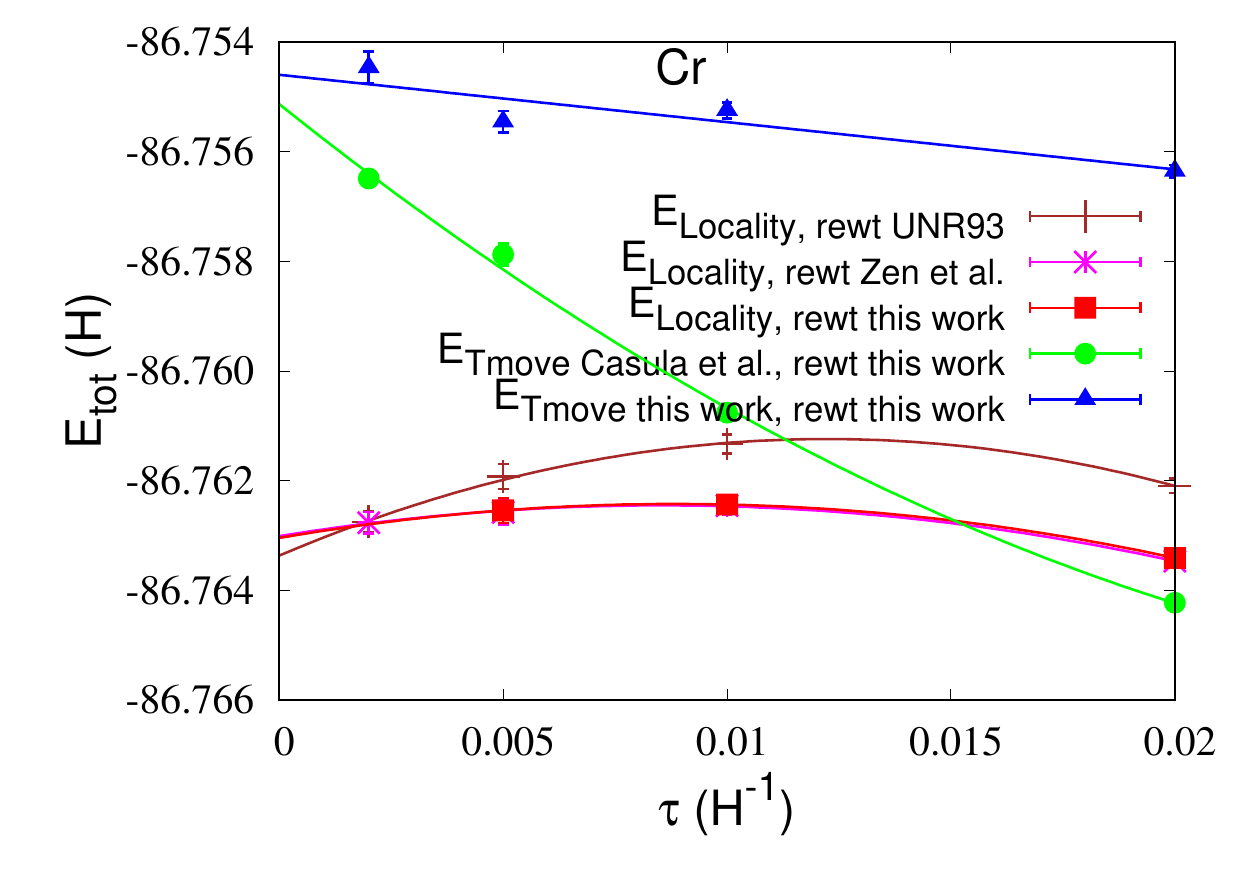}}}\quad
\subfigure[]{{\includegraphics[width=3.5in,height=2.6in,clip]{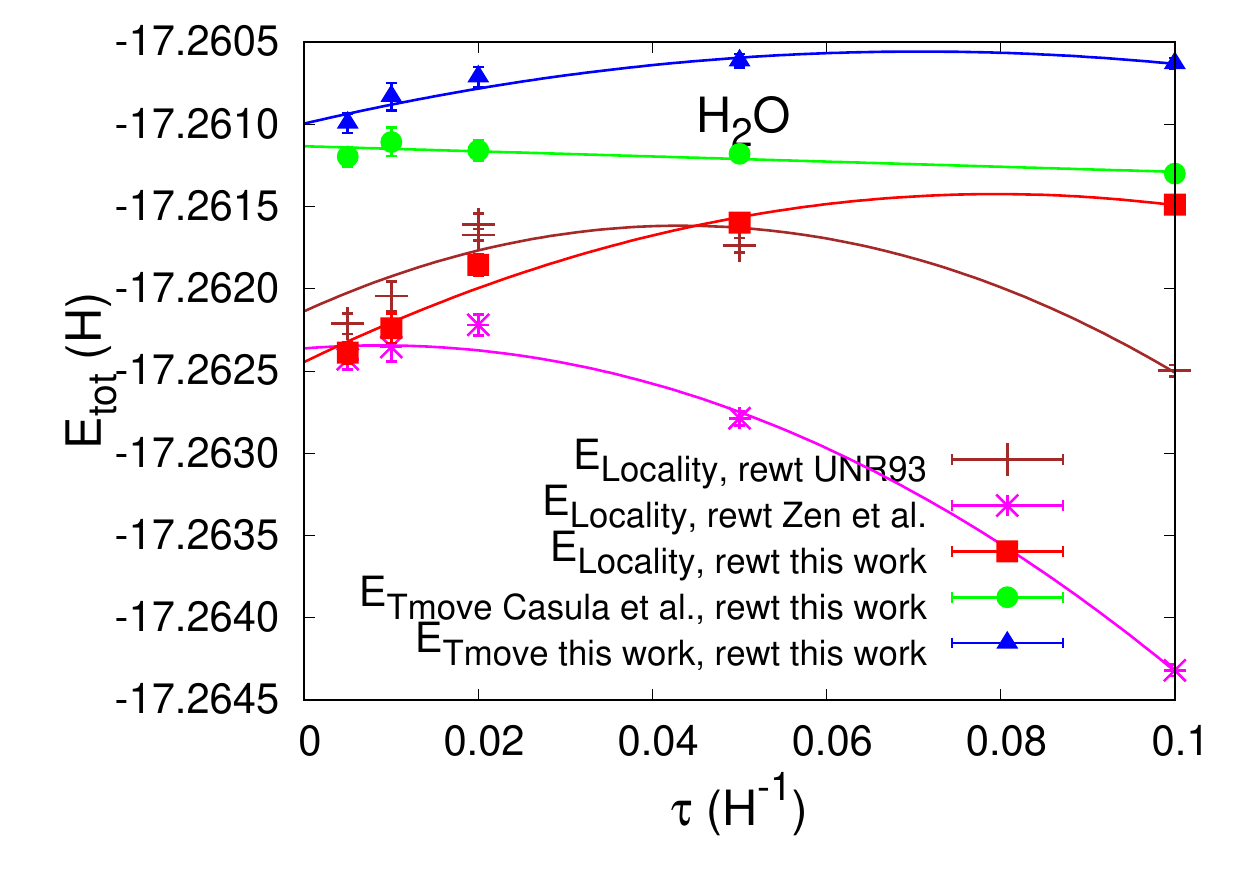}}}\quad
\subfigure[]{{\includegraphics[width=3.5in,height=2.6in,clip]{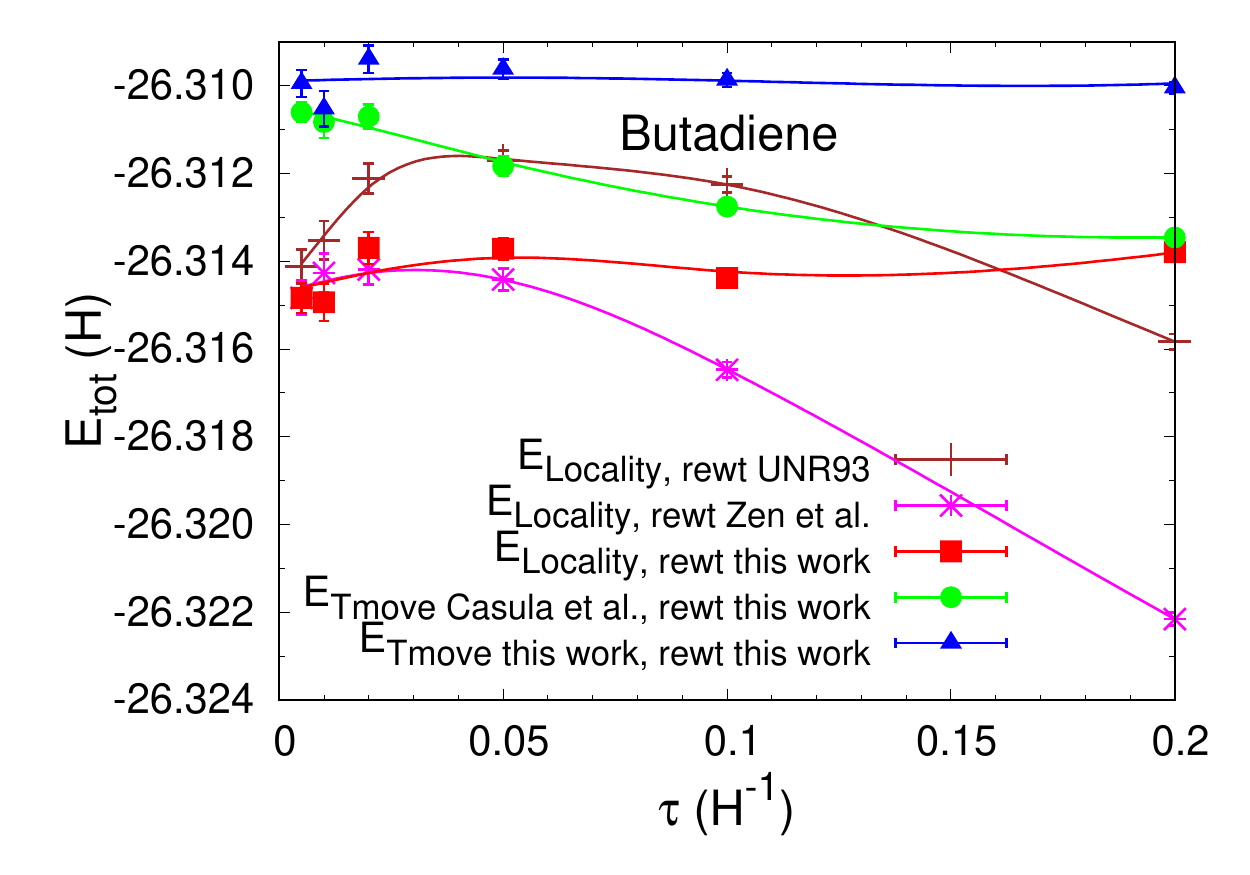}}}\quad
\subfigure[]{{\includegraphics[width=3.5in,height=2.6in,clip]{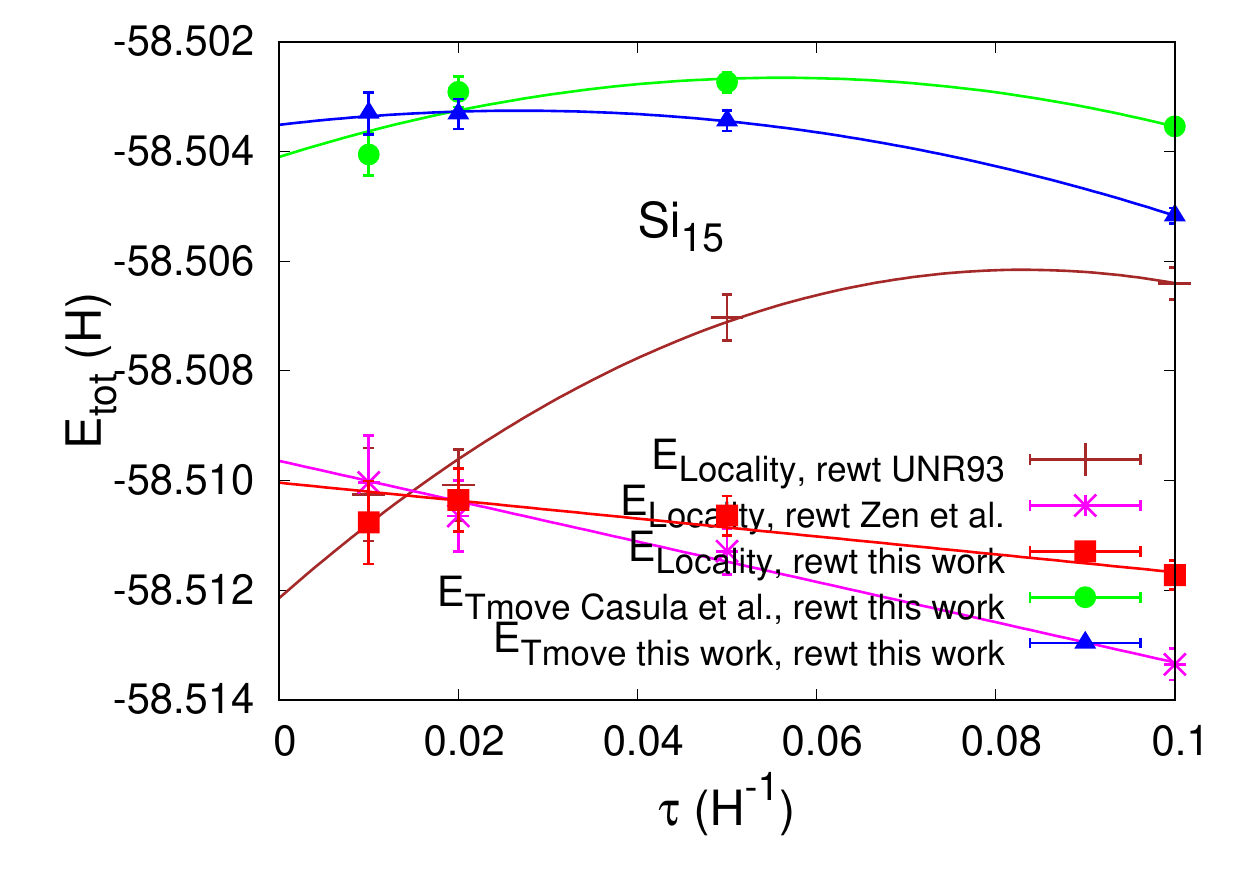}}}\quad
\caption{Time-step errors of the total energy over a smaller range of $\tau$.  The curves labeled ``UNR93" and ``Zen et al." employ the
reweighting factors of Refs.~\onlinecite{UmrNigRun-JCP-93} and \onlinecite{ZenSorGilMicAlf-PRB-16} respectively.
The three locality approximation curves must extrapolate to the same energy at $\tau=0$, as must the two T-moves approximation curves.
The lines are linear or quadratic fits to the data, depending on which gives the smaller estimated error for the $\tau=0$ energy.
Both the reweighting factor and the modified T-moves algorithm proposed in this work contribute to
reducing the time-step errors.}
\label{fig:total_energy_timestep_expan}
\end{figure}

\begin{figure}[b!]
\centering
\subfigure[]{{\includegraphics[width=3.5in,height=2.6in,clip]{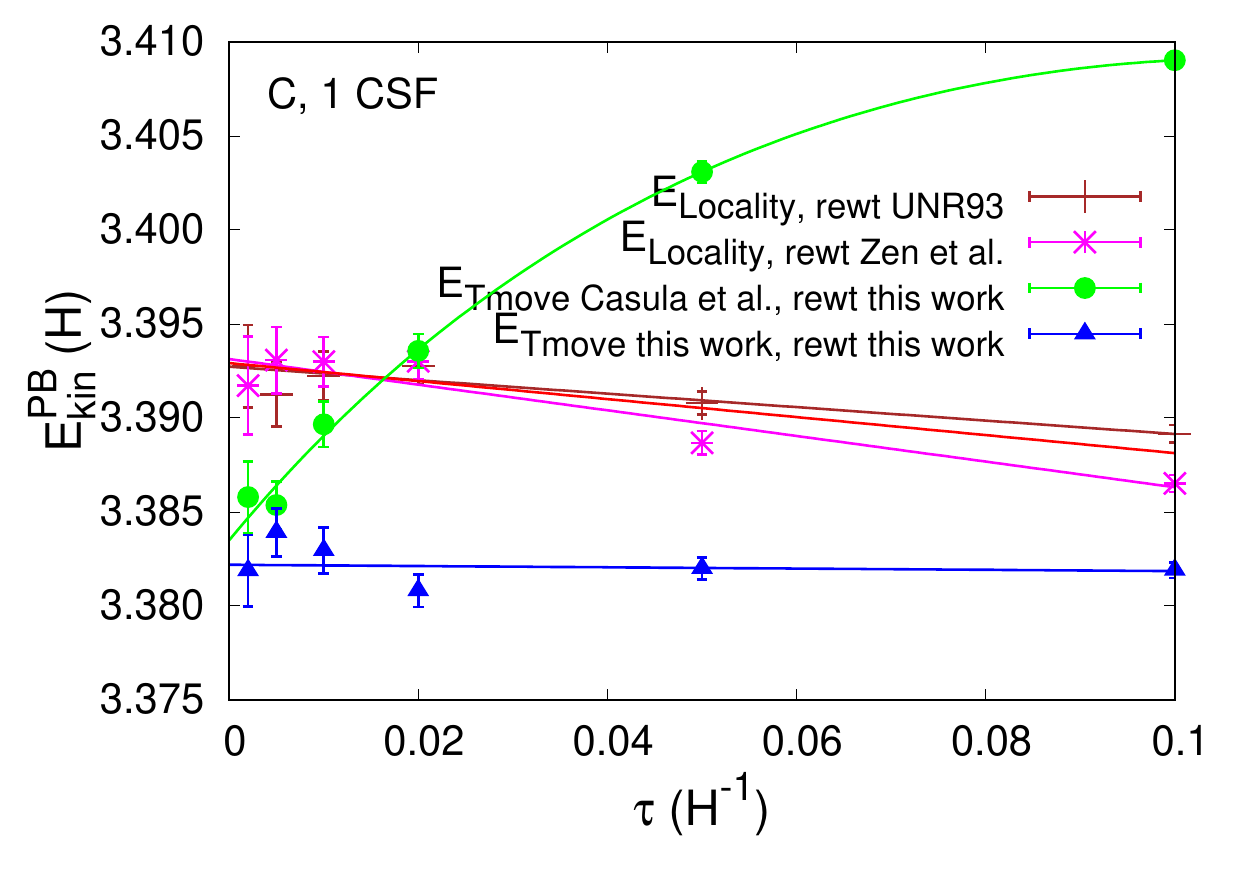}}}\quad
\subfigure[]{{\includegraphics[width=3.5in,height=2.6in,clip]{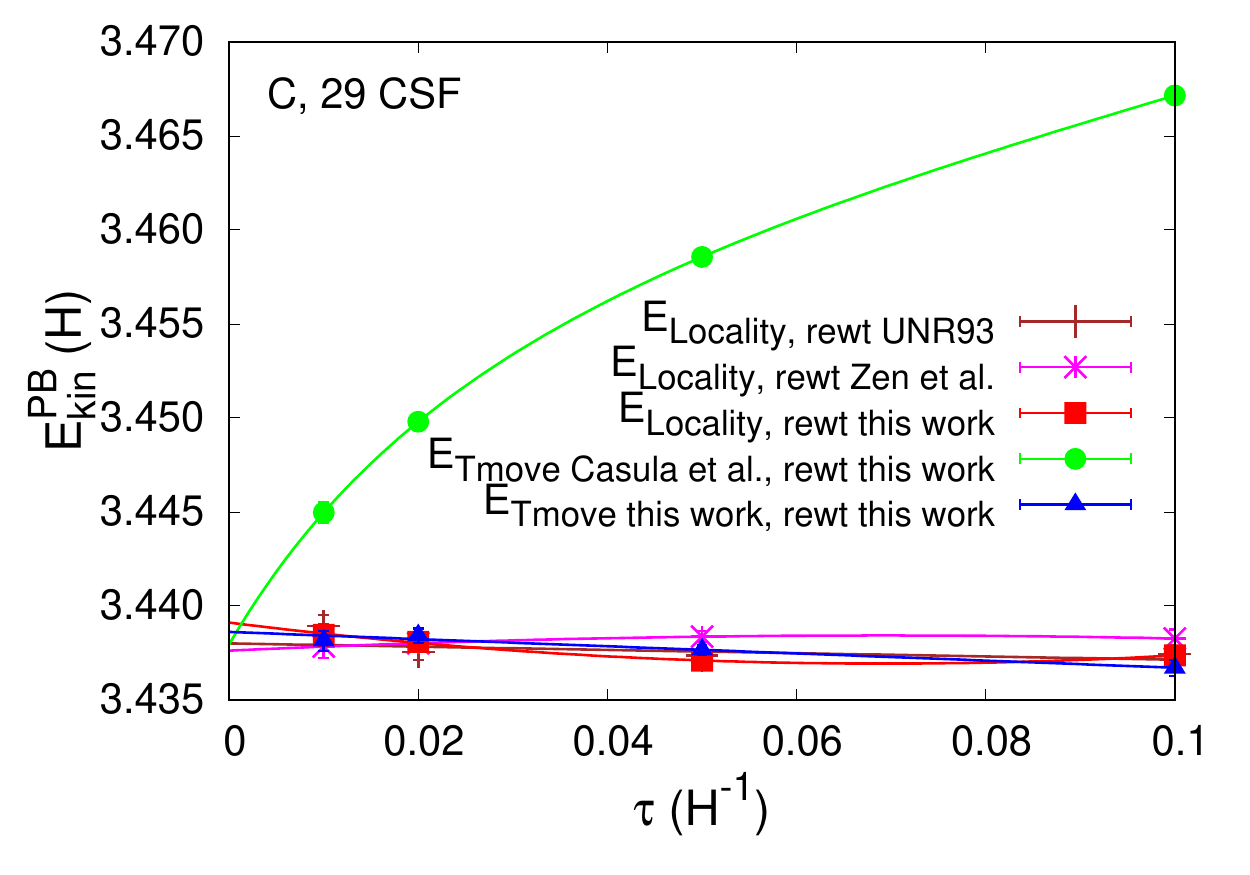}}}\quad\\
\subfigure[]{{\includegraphics[width=3.5in,height=2.6in,clip]{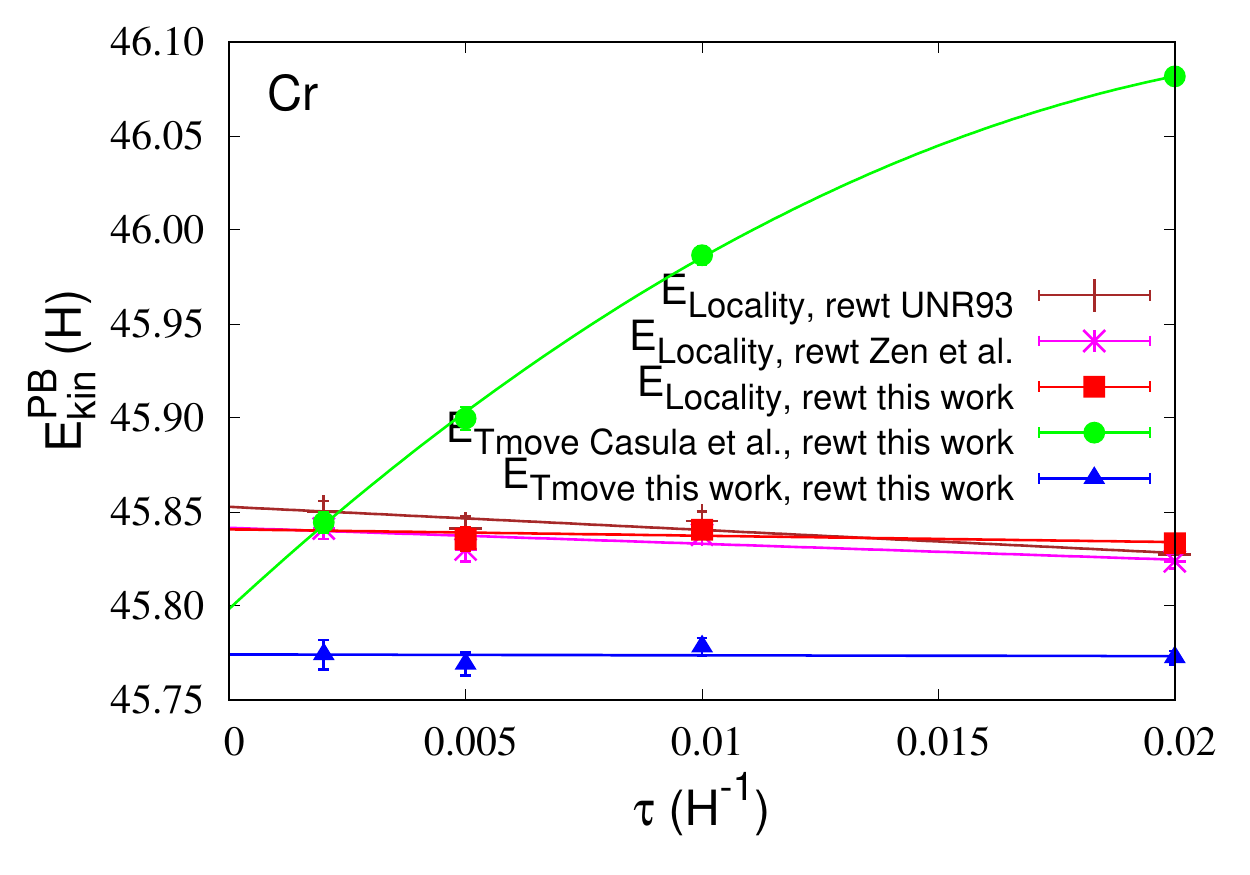}}}\quad
\subfigure[]{{\includegraphics[width=3.5in,height=2.6in,clip]{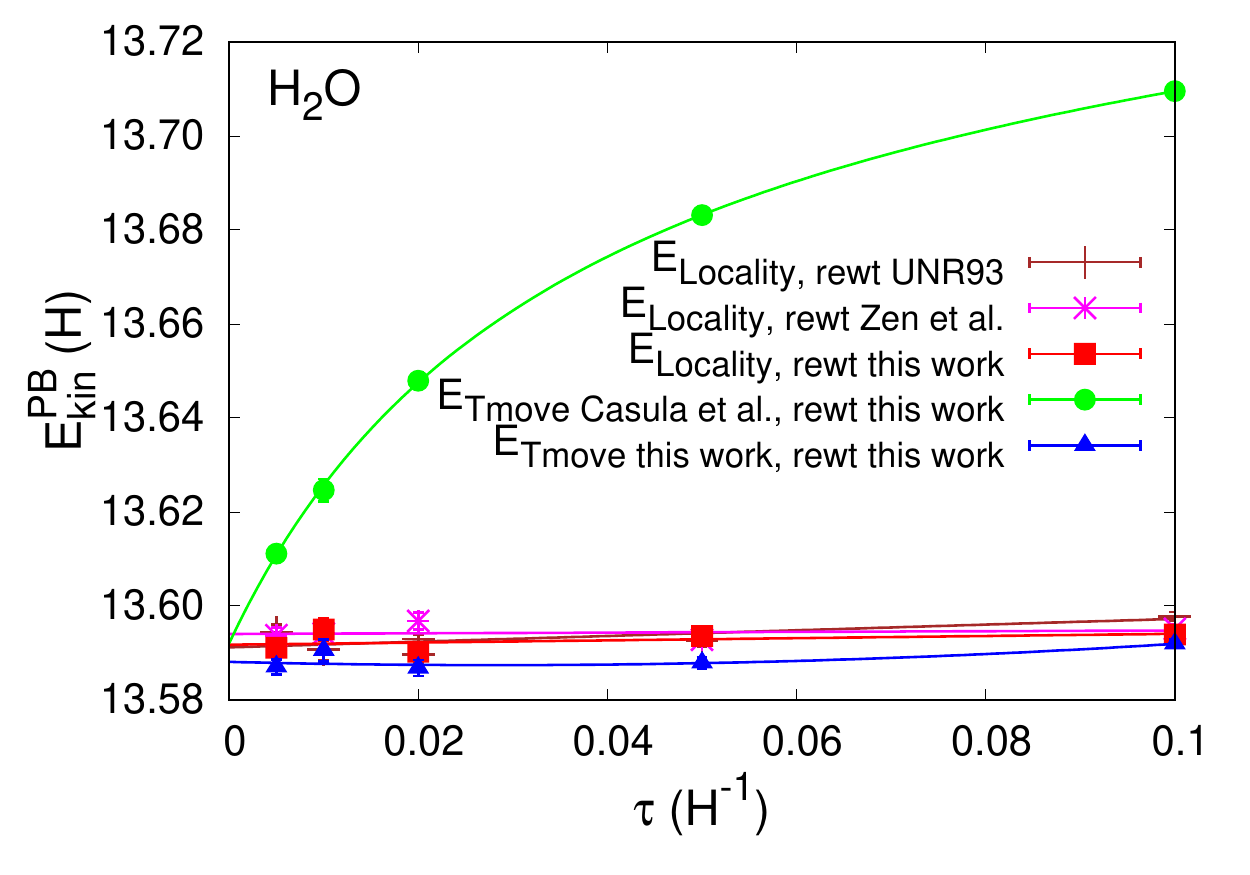}}}\quad\\
\subfigure[]{{\includegraphics[width=3.5in,height=2.4in,clip]{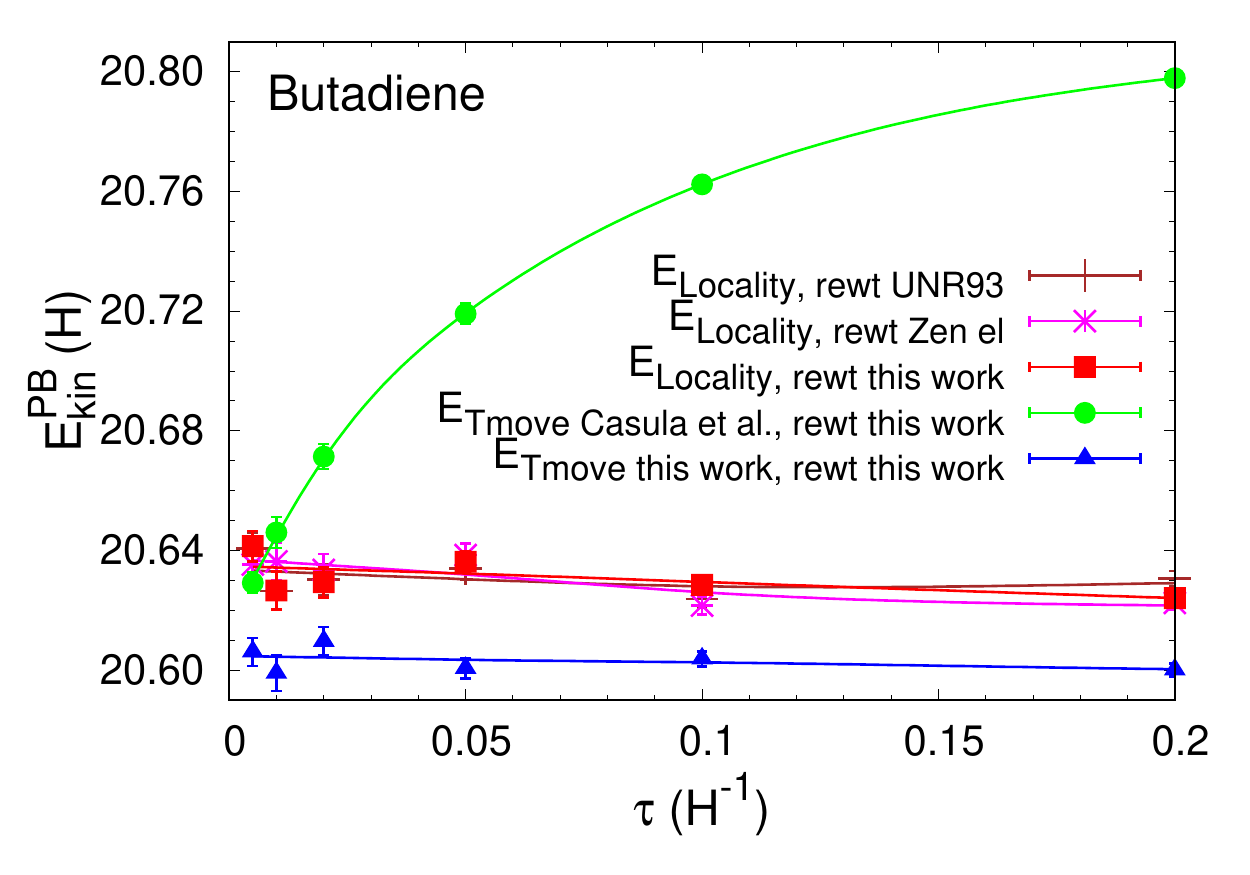}}}\quad
\subfigure[]{{\includegraphics[width=3.5in,height=2.4in,clip]{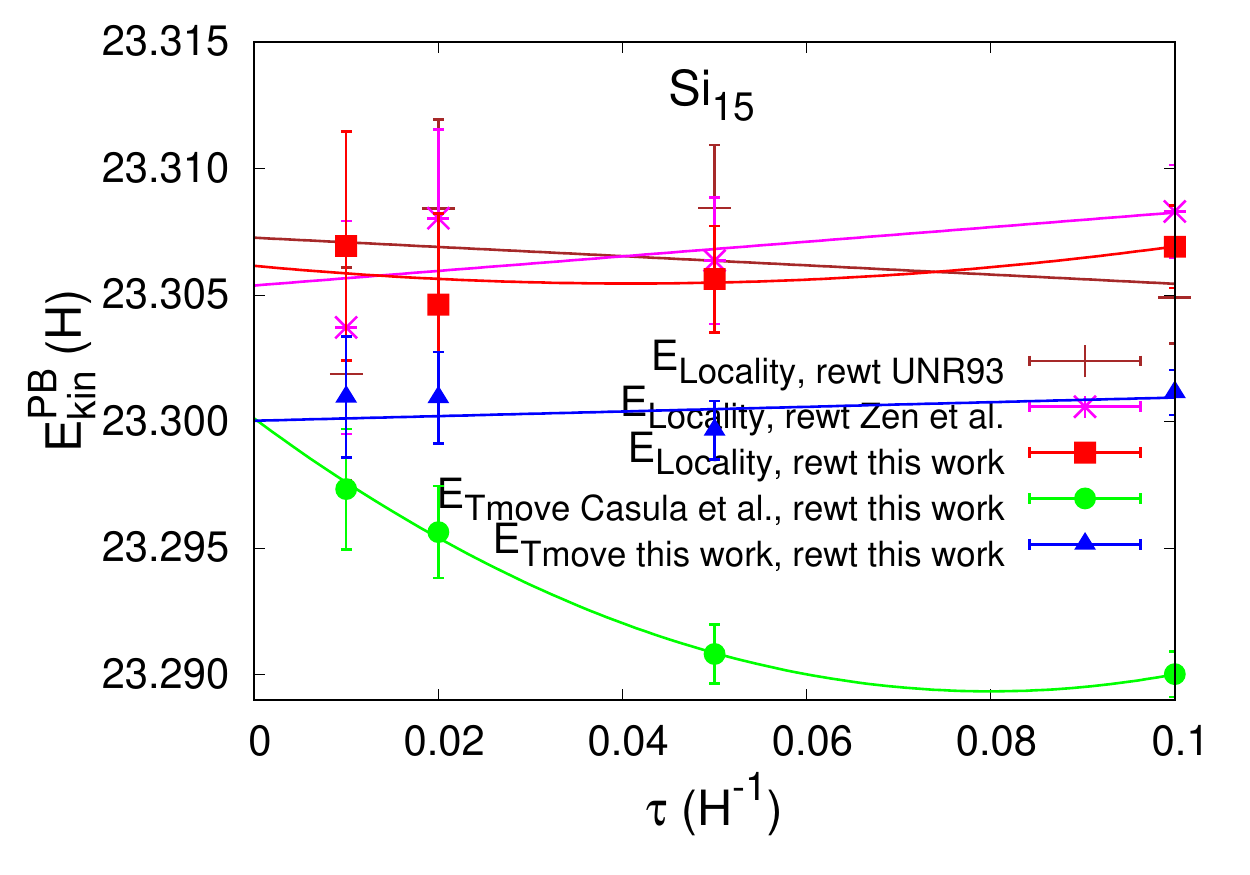}}}\quad
\caption{Same as Fig.~\ref{fig:total_energy_timestep_expan} but for the kinetic energy.
The lines are least squares fits 
to the data using the functional forms described in the text.
Note that the T-moves curves should converge to the same value at $\tau=0$, though this is to some extent obscured by the statistical
errors and the steepness of the kinetic energy curves.}
\label{fig:kinetic_energy_timestep_expan}
\end{figure}

Note that time steps used in Figs.~\ref{fig:total_energy_timestep} and \ref{fig:kinetic_energy_timestep}
(up to 0.5 Ha$^{-1}$ for the five systems with first-row atoms, and up to 0.1 Ha$^{-1}$ for the Cr atom)
are much larger than those commonly used in DMC.
Over this wide range, some of the time-step errors cannot be fit with a low-order polynomial, so the curves shown in these figures
are smoothing splines.
In Figs.~\ref{fig:total_energy_timestep_expan} and \ref{fig:kinetic_energy_timestep_expan}
we show the same data as in Figs.~\ref{fig:total_energy_timestep} and \ref{fig:kinetic_energy_timestep}
but over a time-step range that is only a little larger than that commonly used.
Over this smaller time-step range, we show curves obtained from either a linear or a quadratic fit, choosing
the one that has the smaller estimated error in the extrapolated $\tau=0$ energy.
The only exception to this is the time-step error for the kinetic energy obtained from
Casula et al.'s T-moves approximation.  There, even a quadratic function provides a very poor fit for
some of the systems, so we instead fit to
\beq
E(\tau) &=& E_0 + {E_{1a} \tau \over 1+E_{1b} \tau} + E_2 \tau^2
\eeq
where $E_0$, $E_{1a}$, $E_{1b}$ and $E_2$ are fit parameters.
In each plot we have five curves.  The three locality approximation curves use the reweighting factors of
Ref.~\onlinecite{UmrNigRun-JCP-93}, Ref.~\onlinecite{ZenSorGilMicAlf-PRB-16}, and the reweighting factor advocated
in the present paper.
The two T-moves curves use version 1 of the size-consistent T-moves algorithm of Ref.~\onlinecite{CasMorSorFil-JCP-10},
and the improved T-moves algorithm of the present paper.  We use the reweighting factor of the present paper
for both T-moves curves to ensure that the improvement we demonstrate comes from just the change in the
T-moves algorithm.
For both the total energy and the kinetic energy,
the three locality approximation curves must converge to the same $\tau=0$ value,
as must the two T-moves curves.
This can be seen in the figures
though it is to some extent obscured by
the statistical errors and the steepness of the kinetic energy curves near $\tau=0$ for the T-moves algorithm of Ref.~\onlinecite{CasMorSorFil-JCP-10}.

Comparing the three locality approximation curves in each sub-plot of Fig.~\ref{fig:total_energy_timestep} we see that
the time-step error is considerably reduced upon using the reweighting
factor of Eq.~\ref{eq:rewt} rather than those in Refs.~\onlinecite{UmrNigRun-JCP-93} or \onlinecite{ZenSorGilMicAlf-PRB-16}.
In particular the Ref.~\onlinecite{UmrNigRun-JCP-93} and \onlinecite{ZenSorGilMicAlf-PRB-16} reweightings typically give a large negative
time-step bias in the total energy at large $\tau$.
Further, the Ref.~\onlinecite{UmrNigRun-JCP-93} reweighting, when used with the locality approximation, often
has a positive hump for $\tau$ in the range $(0,0.1)$ Ha$^{-1}$ which makes it difficult to
extrapolate to $\tau=0$.  This hump gets bigger with increasing system size and results in a non-size-consistent
time-step error as noted in Ref.~\onlinecite{ZenSorGilMicAlf-PRB-16}.
It is much reduced or eliminated upon using the new reweighting factor, and almost completely disappears upon
going from the locality approximation to our improved T-moves approximation.
We note that our goal in this paper is to choose a reweight factor that has a small time-step error
for all systems, as opposed to one that achieves a good cancellation of time-step errors for
a molecule and its fragments.  In appendix~\ref{app:size_consis_rewt} we present a choice
that attempts to achieve the latter goal.

Comparing the (a) and (b) subplots in Figs.~\ref{fig:total_energy_timestep}, \ref{fig:kinetic_energy_timestep},
~\ref{fig:total_energy_timestep_expan} and \ref{fig:kinetic_energy_timestep_expan},
we see that the $\tau=0$ limits for the locality approximation and the T-moves approximation curves
get closer upon using the more accurate wave function with 29 CSFs.
The difference between the extrapolated total energies is 1 mHa for 1 CSF but only 0.1 mHa for 29 CSFs.
This of course is as expected since in the $\PsiT=\Psiz$ limit both approximations become exact at $\tau=0$ for all observables.
Note however, that the time-step error of the kinetic energy from the T-moves approximation of Ref.~\onlinecite{CasMorSorFil-JCP-10}
at finite $\tau$ is actually a little larger for the wave function with 29 CSFs than the one with 1 CSF.
This is because this approximation fails to sample $\Psizsq$ in the $\PsiT \to \Psiz$ limit, and so expectation values of
operators that do not commute with the Hamiltonian are not bias free in this limit, except at $\tau=0$.

Fig.~\ref{fig:kinetic_energy_timestep} shows that for most of the systems tested, the T-moves approximation of Ref.~\onlinecite{CasMorSorFil-JCP-10}
has a very steep (positive or negative) time-step dependence for the kinetic energy at small $\tau$ which makes it very difficult
to extrapolate to $\tau=0$.  This is cured by our improved T-moves approximation.
The difference between the two T-moves approximations is smallest for Si$_{15}$ because the Si pseudopotential
is not strongly nonlocal, i.e., the difference between the s and p angular momentum components
is smaller than for carbon or oxygen. (See Fig. 2 in the Supplementary Material~\cite{supplementary}.)  The fact that the d component
differs considerably from the s and p components in silicon is not very important since there are no occupied d orbitals.

Note that, for all the systems tested, the curves for the reweighting factor and the T-moves of this work
have a simple shape and are rather flat.
The other combinations of reweighting factor and locality approximation or T-moves may have
these desirable features for some systems but not for all of the ones tested.
Hence it appears that the reweighting factor and the T-moves of this work enable one to use
time-steps about an order of magnitude larger than those commonly used, particularly if
one is interested in expectation values of observables that do not commute with the energy,
such as the kinetic energy.

\subsection{Efficiency of T-moves}
We now demonstrate that the T-moves approximations are more efficient than the locality approximation
even though the time per Monte Carlo (MC) step is a bit larger when $\tau$ is large.
The statistical error of the energy for a fixed number of MC steps is proportional to
$\sigma_E \sqrt{\Tcorr}$ where $\sigma_E$ is the root-mean-square fluctuation of the energy and
$\Tcorr$ is the autocorrelation time in units of the number of MC steps.
Hence the efficiency measure is $1/(T \Tcorr \sigma_E^2)$, where $T$ is the computer time
per MC step.
Typically $\sigma_E$ and $T$ are nearly independent of $\tau$ at small $\tau$, whereas
$\Tcorr \propto 1/\tau$ at small $\tau$.

Table~\ref{tab:efficiency} demonstrates the higher efficiency of the two T-moves approximations
relative to the locality approximation.
In order to study just the effect of the modifications to the T-moves algorithm,
each of the three calculations uses the reweighting factor of Eq.~\ref{eq:rewt}.
The main observation to be gleaned from Table~\ref{tab:efficiency} is that
the additional electron moves generated by the T-moves serve to decorrelate the
the electron positions and reduce $\Tcorr$.
The T-moves proposed in this paper tend to have a slightly smaller $\sigma_E$ than
the T-moves of Casula et al., probably because the former correctly samples $\vert \PsiT \vert^2$
in the $\PsiT \to \Psiz$ limit and therefore has a smaller density of electrons near
the nodal surface of $\PsiT$.  However, it has a slightly larger $\Tcorr$ because
of the rejection of the proposed T-moves, and the two effects tend to cancel when
computing the efficiency.

At first sight it seems surprising that the computer time does not increase more than
it does going from the locality approximation to Casula et al. T-moves to the
T-moves of this work.
The reason why the time for Casula et al. T-moves is not considerably more than for
the locality approximation is that the values of the wave function on the spherical quadrature points
calculated when evaluating the local energy are saved and reused when doing the T-moves.
It is only when an electron makes an off-diagonal T-move that the wave function needs to be reevaluated when making T-moves
for the remaining electrons of that walker.
However, most of the time, especially at small $\tau$, the T-moves are diagonal.
This is also the reason why the computer time for our T-moves approximation is
virtually the same as that for the Casula et al. T-moves approximation -- the
wave function values for the reverse Metropolis-Hastings move are needed only
if an off-diagonal T-move is made.
Moreover, as discussed earlier, one could use the same quadrature grid for forward
and reverse moves.  
However, despite using an independent grid for the reverse moves in Table~\ref{tab:efficiency}, the computer times for the two T-moves approximations are nearly the same.
Finally note that in the $\tau \to 0$ limit, the computer times for all three methods
will be equal.

\begin{table*}[htb]
\caption{Comparison of efficiency, $1/(T \Tcorr \sigma_E^2)$, of locality approximation, Casula et al. T-moves and the
T-moves of this paper, for Cr and butadiene, each at the smallest and largest $\tau$ used.
The computer time $T$ is relative to the time for the locality approximation.
The T-moves approximations are more efficient than the locality approximation despite taking slightly more time per MC move.
}
\label{tab:efficiency}
\begin{tabular}{l|d|d|d}
\hline
\multicolumn{1}{c|}{Quantity} & \multicolumn{1}{c|}{locality} & \multicolumn{1}{c|}{Casula T-moves} & \multicolumn{1}{c}{this work T-moves} \\
\hline
\multicolumn{4}{c}{Cr, $\tau=0.002$ Ha$^{-1}$} \\
\hline
$T$ & 1 & 1.01 & 1.01 \\
$\sigma_E$ & 1.24 & 1.22 & 1.21 \\
$\Tcorr$ &   28.0 &         20.8 &          21.2 \\
efficiency & 0.023 &        0.032 &         0.032 \\
\hline

\multicolumn{4}{c}{Cr, $\tau=0.1$ Ha$^{-1}$} \\
\hline
$T$        & 1 &           1.15 &           1.15 \\
$\sigma_E$   & 1.26 &         1.30 &          1.22 \\
$\Tcorr$   & 1.69 &         1.29 &          1.31 \\
efficiency & 0.37 &         0.40 &          0.45 \\
\hline
\multicolumn{4}{c}{Butadiene, $\tau=0.005$ Ha$^{-1}$} \\
\hline
$T$ & 1 & 1 & 1 \\
$\sigma_E$ & 0.62 & 0.61 & 0.60 \\
$\Tcorr$ &   53.7  &        48.1  &         43.4  \\
efficiency & 0.048 &       0.056  &         0.064 \\
\hline

\multicolumn{4}{c}{Butadiene, $\tau=0.5$ Ha$^{-1}$} \\
\hline
$T$        &  1 &           1.07 &          1.07 \\
$\sigma_E$   & 0.67 &         0.64 &          0.62 \\
$\Tcorr$   & 1.97 &         1.39 &          1.48 \\
efficiency & 1.13 &         1.64 &          1.64 \\
\hline
\end{tabular}
\end{table*}

\section{Outlook}
\label{sec:outlook}
We have demonstrated that by introducing a Metropolis-Hastings accept-reject step after each
one-electron T-move and by using a modified reweighting factor, it is possible to use an order of
magnitude larger time step than is commonly used in DMC.  In particular,
the time-step error of our modified T-moves algorithm is nearly linear with
a small slope for the first-row systems we studied in the range $(0,0.2)$ Ha$^{-1}$,
and for the Cr atom in the range $(0,0.05)$ Ha$^{-1}$.
Preliminary studies on model systems indicate that by using modified position-dependent
drift and diffusion terms in the Green's function, further reductions in the time-step error are possible.
The extension of these ideas to real systems will be the subject of a future paper.

\section{Supplementary Material}
In the supplementary material we demonstrate the weak dependence of the time-step error on the parameter $a$ in Eq.~\ref{eq:vbar} for H$_2$O and demonstrate that the reweighting factor proposed in this paper reduces the time-step error of the total energy
not only when the T-moves approximation is used but also when the locality approximation is used.
We also discuss the degree of nonlocality in the carbon, oxygen, and silicon pseudopotentials.
Finally, we provide the VMC energies for each system to indicate the quality of the trial wave functions used.

\section{Data Availability}
The data that support the findings of this study are available within the article, the supplementary material of
the arXiv version of this paper~\cite{AndUmr-ARX-21} and from the corresponding authors.

\acknowledgements
We thank Michele Casula and Claudia Filippi for discussions and comments on the manuscript, and
Andrea Zen and Dario Alfe for discussions about size consistency.
The computations on Si$_{15}$ were performed on the Bridges computer at the Pittsburgh Supercomputing Center supported by NSF grant ACI-1445606, as part of the XSEDE program supported by NSF grant ACI-1548562.
This work was supported in part by the AFOSR under grant FA9550-18-1-0095.

\appendix
\section{Average velocity}
\label{app:integ_vel}

The velocity $\Vvec$ diverges near a node.  When the initial position of a walker is near
a node, it is a very poor approximation to assume that the velocity is constant during
time step $\tau$ as is implicitly done in Eq.~\ref{eq:Green2}.
A better approximation can be derived by making a simple ansatz for the form of the
wave function near the node and integrating the velocity over $\tau$.
Keeping all except one electron fixed, here we make the following ansatz for the wave function:
\beq
\psi(\rvecp) &=& \psi(\rvec) \left(1 + a\vvec\cdot(\rvecp-\rvec)\right)^{1/a} \\
\mbox{So, \hskip 3mm} \nablavec_{\rvecp}\psi(\rvecp) &=& \psi(\rvec) \left(1 + a\vvec\cdot(\rvecp-\rvec)\right)^{1/a-1}\vvec.
\eeq
Let $x=|\rvecp-\rvec|$ be the drift distance, then
\beq
{dx \over d\tau} \;=\; {|\nablavec_{\rvecp}\psi(\rvecp)| \over \psi(\rvecp)}
&=& {v \over 1 + a v x}.
\eeq
Integrating and solving for $x$,
\beq
\label{eq:drift}
x &=& {-1 + \sqrt{1+2av^2\tau} \over av}.
\eeq
So, the average velocity over $\tau$ is
\beq
\label{eq:vvecbar}
\bar{\vvec} &=& {-1 + \sqrt{1+2av^2\tau} \over av^2 \tau} \vvec \;=\;
\left\{\begin{array}{lll} \displaystyle
\left(1-{av^2\tau \over 2}\right) \vvec & \quad av^2\tau \to 0 \\
\sqrt{2 \over a \tau} \; \hat{v}   & \quad av^2\tau \to \infty
\end{array}
\right. .
\eeq
This is Eq.~35 of Ref.~\onlinecite{UmrNigRun-JCP-93}.
For $a=1$ the local behavior of the wave function is linear, for $a<1$ it is convex,
and for $a>1$ it is concave.
The time-step error is a fairly weak function of $a$, for $a \in [0.1,1.5]$ and
the optimal value depends on the system and other details of the algorithm.
In this work we use $a=0.5$, which is reasonable for finite systems, which have convex tails.
In the Supplementary Material~\cite{supplementary} we compare the time-step errors for H$_2$O for $a=0.5$ and $a=0.25$.
Of course many equally reasonable ansatze can be made for the wave function and they result in
average velocities that differ in detail but have the general property that they decay
with increasing $\tau$.
In a future paper, we plan to explore further reductions in the time-step error that can be achieved
by combining a physically motivated position dependent $a$ with a modified formula for the diffusion.
Tests on model systems show promising results.

\section{T-moves Energy Bound}
\label{app:energy_bound}
One can show that the ground state energy in the T-moves approximation is strictly greater than the ground state energy in the locality
approximation. Let $H$ be the exact Hamiltonian, and consider an arbitrary basis $\{ | B_i \rangle \}$ so that $H_{ij} = \langle B_i | H |
B_j \rangle$. For a trial wave function $\Psi$, the Hamiltonian in the locality approximation is defined as
\begin{equation}
H^{LA}_{ii} = H_{ii} + \sum_{j} H_{ij}\frac{\Psi_j}{\Psi_i}.
\end{equation}
The Hamiltonian in the T-moves approximation has nonzero matrix elements both on and off the diagonal. The diagonal terms are given by
\begin{equation}
H^{T}_{ii} = H_{ii} + \sum_{j} H_{ij}\frac{\Psi_j}{\Psi_i},
\end{equation}
where the sum is taken only over $i$ and $j$ such that $H_{ij}\frac{\Psi_j}{\Psi_i} \geq 0$. The off-diagonal terms are given by
\begin{equation}
H^{T}_{ij} = H_{ij}
\end{equation}
for $i$ and $j$ such that $H_{ij}\frac{\Psi_j}{\Psi_i} < 0$.

We now show that, for an arbitrary state $| \Phi \rangle =
\sum_i c_i | B_i \rangle$, we have $\langle \Phi | H^T | \Phi \rangle \geq \langle \Phi | H^{LA} | \Phi \rangle$.
The only difference between $H^T$ and $H^{LA}$ is their treatment of matrix elements for which $H_{ij}\frac{\Psi_j}{\Psi_i} < 0$. It follows that
\begin{equation}
\langle \Phi | H^T - H^{LA} | \Phi \rangle = \sum_{i,j} \left( c_iH_{ij}c_j - c_i H_{ij}\frac{\Psi_j}{\Psi_i} c_i \right),
\end{equation}
where the sum is only over indices $i$ and $j$ such that $H_{ij}\frac{\Psi_j}{\Psi_i} < 0$. Because $H_{ij}\frac{\Psi_j}{\Psi_i} < 0$  implies
$H_{ji}\frac{\Psi_i}{\Psi_j} < 0 $, this sum can be symmetrized:
\begin{equation}
\label{bound_symm}
\sum_{i,j} \left( c_iH_{ij}c_j - c_i H_{ij}\frac{\Psi_j}{\Psi_i} c_i \right) = \frac{1}{2} \sum_{i,j} \left( 2 H_{ij} c_ic_j - c^2_i
H_{ij}\frac{\Psi_j}{\Psi_i} - c^2_j H_{ij}\frac{\Psi_i}{\Psi_j}\right),
\end{equation}
using $H_{ij} = H_{ji}$. Because $H_{ij}\frac{\Psi_j}{\Psi_i} < 0$, $H_{ij}\frac{\Psi_j}{\Psi_i} = - |H_{ij}| \left| \frac{\Psi_j}{\Psi_i}
\right|$. Eq.~\ref{bound_symm} then reduces to
\begin{equation}
 \frac{1}{2} \sum_{i,j} |H_{ij}| \left( 2 \textrm{sgn}(H_{ij}) c_i c_j + c^2_i \left| \frac{\Psi_j}{\Psi_i} \right| + c^2_j \left|
 \frac{\Psi_i}{\Psi_j} \right| \right).
\end{equation}
Noting that the term in parenthesis may be written as the square of a real number, our final result is
\begin{equation}
\langle \Phi | H^T - H^{LA} | \Phi \rangle =
\frac{1}{2} \sum_{i,j} |H_{ij}| \left(c_i \sqrt{\left| \frac{\Psi_j}{\Psi_i} \right|} + \textrm{sgn}(H_{ij}) c_j \sqrt{\left| \frac{\Psi_i}{\Psi_j}
\right|}\right)^2,
\end{equation}
which is clearly positive. It follows that the ground state energy of $H^T$ is greater than the ground state energy of $H^{LA}$.

\section{Alternative Reweighting Factor}
\label{app:size_consis_rewt}

The reweighting factor described in this paper gives a small time-step error even for large systems
(see Fig.~\ref{fig:total_energy_timestep}), but it is not manifestly size consistent.
In this appendix, we discuss a reweighting factor that ensures size consistency, but has a different drawback.

Consider a system, $AB$ consisting of two widely separated fragments, $A$ and $B$ and a wave function
which factors, i.e., $\Psi_{AB} = \Psi_A \Psi_B$.
Then as discussed in Ref.~\onlinecite{ZenSorGilMicAlf-PRB-16}, the energy is size consistent, i.e., $E_{AB} = E_A + E_B$ if $\Sbar_{AB}(\Rvec_A,\Rvec_B) = \Sbar_A(\Rvec_A) + \Sbar_B(\Rvec_B)$, which is
not the case for Eq.~\ref{eq:Sbar}.
In the current DMC algorithm, the reweighting factor is calculated only once per Monte Carlo step, after all electron positions have been updated.
Instead, one can calculate separate contributions to the reweighting factor after each one-electron move.
In particular, define the one-electron energy of electron $i$ as
\beq
\label{eq:Eloci}
E^{i}_{L}(\Rvec) =
\frac{1}{\PsiT(\Rvec)}\left(\left(-{1 \over 2} \nabla_i^2 + v_{\rm L}(\rvec_i) + \frac{1}{2}\sum_{j}{1 \over r_{ij}} \right) \PsiT(\Rvec)
+ \int d\rvecp_i \; v_{\rm NL}(\rvec_i,\rvecp_i) \PsiT(\rvec_1,\cdots,\rvecp_i,\cdots,\rvec_N) \right), \nonumber \\
\eeq
where $v_{L}$ and $v_{NL}$ are the local and nonlocal parts of the pseudopotential, respectively.
Then one can define an alternative reweighting factor $\bar S^{(2)}(\Rvec)$ analogous to Eq.~$\ref{eq:Sbar}$, with the local energy $E_L(\Rvec)$ replaced with the one-electron local energy $E^{i}_{L}(\Rvec)$ and the $V^{2}/N$ replaced by $v^{2}_i$:
\beq
\label{eq:Sbar2}
\bar S^{(2)}(\Rvec) = \ET-\Eest + \sum^{N}_i {E^{i}_{\rm cut}(\Rvec) \over 1 + (v_{i}^2\tau)^2},
\eeq
with
\beq
\label{eq:Ecuti}
E^{i}_{\rm cut}(\Rvec) = \min\left\{\left|\frac{\Eest}{N}-E^{i}_{\rm L}(\Rvec)\right|,10 \sigma_E\right\} \sgn\left(\frac{\Eest}{N}-E^{i}_{\rm L}\right).
\eeq
This choice is size consistent and does not add to the computational cost since only three components of the Laplacian must be calculated per one-electron move.
However this reweight factor has the disadvantage that it does not have a zero-variance principle - as $\PsiT$ approaches the true ground state, the reweighting factor does not have zero variance.
(Recall that the reweighting factor in the standard DMC algorithm is 1, for $\PsiT=\Psiz$.)
This is the reason why we do not use this choice of the reweight factor for the calculations in this paper.

\bibliographystyle{apsrev4-1}
\bibliography{umrigar,sorella,biblio,mitas,needs,qmc,nonloc_paper}

\end{document}